\documentclass[a4paper,fleqn,usenatbib]{mnras}

\usepackage[T1]{fontenc}
\usepackage{ae,aecompl}

\usepackage{graphicx}
\usepackage{amsmath}
\usepackage{stmaryrd}

\newcommand{\simnot}{\mathord{\sim}}
\mathchardef\mhyphen="2D

\title[Spatially-Offset AGN in CLASS]{Spatially-Offset AGN Candidates in the CLASS Survey}

\author[C.~J.~Skipper and I.~W. A.~Browne]{
Chris~J.~Skipper$^1$
and Ian~W.A.~Browne$^1$
\\
$^{1}$Jodrell Bank Centre for Astrophysics, Alan Turing Building, The University of Manchester, Manchester, M13 9PL, UK
}

\date{Accepted XXX. Received YYY; in original form ZZZ}

\pubyear{2018}

\begin{document}
\label{firstpage}
\pagerange{\pageref{firstpage}--\pageref{lastpage}}

\maketitle

\begin{abstract}
Prompted by a recent claim by Barrows et al. that X-ray AGN are often found significantly offset from the centres of their host galaxies, we have looked for examples of compact radio sources which are offset from the optical centroids of nearby ($z < 0.2$) galaxies. We have selected a sample of 345 galaxies from the Sloan Digital Sky Survey (SDSS) galaxy catalog which have nearby compact radio sources listed in the Cosmic-Lens All Sky Survey (CLASS) catalog. We find only three matches ($\simnot 0.87$ per cent of the sample) with offsets greater than 600~milliarcsec (mas), which is considerably fewer than we would have expected from the Barrows et al. X-ray survey. We fit our histogram of offsets with a Rayleigh distribution with $\sigma = 60.5$~mas, but find that there is an excess of objects with separations greater than $\rm\simnot 150~mas$. Assuming that this excess represents AGN with real offsets, we place an upper limit of $\simnot 17$ per cent on the fraction of offset AGN in our radio-selected sample. We select 38 objects with offsets greater than 150~mas, and find they have some diverse properties: some are well known, such as Mrk 273 and Arp 220, and others have dust lanes which may have affected the optical astrometry, while a few are strong new candidates for offset AGN.
\end{abstract}




\section{Introduction}

\subsection{Offset AGN}

It is now widely accepted that the majority of galaxies host a supermassive black hole (SMBH) at their centre, and the accretion of matter into SMBHs accounts for the high bolometric luminosities of active galactic nuclei (AGN). It therefore follows that merging or interacting galaxies may contain more than one SMBH, also in some stage of merger or interaction. It has been predicted since the early 1980s \citep[e.g.][]{Begelman1980, Roos1981} that SMBHs in interacting galaxies may form massive binary systems, which may eventually coalesce in a burst of gravitational waves \citep{Thorne1976}.

The offset of a SMBH from the centre of its host galaxy can result from a galaxy merger if, for example, a binary pair of SMBHs forms from the remains of the merger components, or a SMBH is left exposed after its host galaxy was tidally shredded in a merger \citep{Condon2017}, or alternatively if the anisotropic emission of gravitational waves from a coalesced binary pair results in a gravitational 'kick' that dislodges the SMBH from the galaxy centre \cite[e.g.][]{Peres1962, Bekenstein1973, Comerford2014}. If the SMBH is accreting gas that has been disrupted in the merger event then it will emit strongly across the electromagnetic spectrum, and can be detected as an AGN displaced from the centre of mass of its host galaxy. For convenience, we will refer to these as "offset AGN" throughout this paper.

Offset AGN can be detected in several ways. Firstly, offset-AGN candidates can be identified by spectroscopic searches, which look for emission lines that are offset in velocity space from the host galaxy \citep[e.g.][]{Barrows2016} or are double-peaked \citep[e.g.][]{Rubinur2017,Tytler1985}. However, some follow-up surveys \citep[e.g.][]{Tingay2011,Gabanyi2014} have failed to find any evidence of dual AGN in galaxies with double-peaked narrow-line emission.

Another way to identify offset-AGN candidates is to search for sources that are spatially offset from the optical centre of the galaxy using either X-ray \citep[e.g.][]{Comerford2015,Barrows2016,Koss2011,Liu2013} or radio \citep[e.g.][]{Fu2011} imaging. This method can identify candidate offset AGN, but cannot provide confirmation since the centre of optical light may not be a true measurement of the centre of the galaxy in cases where there are dust lanes or regions of strong star formation. A related technique was employed by \cite{Lena2014}, who searched for offsets between the nuclear optical (or near infrared) point source and the mean photocentre of 14 nearby elliptical galaxies. Finally, a third way is to search specifically for binary AGN using high-resolution radio maps \citep[e.g.][]{Burke-Spolaor2011,Deane2014}.

Although most techniques for identifying offset or binary AGN have found promising candidates, few have produced confirmed examples. A recent paper by \cite{Barrows2016} has claimed a high rate of success by searching for X-ray-loud offset AGN in a sample of Sloan Digital Sky Survey (SDSS) selected galaxies, by cross-matching the galaxy positions with observations in the \textit{Chandra} archive. The authors identify X-ray emission from the AGN by requiring that the 2-10~keV luminosity exceeds $\rm 10^{42}~erg~s^{-1}$, and the hardness ratio, defined such that $HR = (H - S)/(H + S)$, is greater than or equal to -0.1. They find that, out of a parent sample of 48 AGN with \textit{Chandra} X-ray detections, there are 18 with the X-ray source significantly offset from the stellar core, which suggests that offset AGN are more prevalent than expected.

In contrast, radio astronomers have traditionally relied upon the assumption that radio AGN are always coincident with their host galaxies, which has led to the identification of tens of thousands of AGN. However, it is possible that some real examples of displaced, radio-loud AGN could have been overlooked by too rigorous a reliance upon this assumption. To investigate if real radio AGN offset might have been missed in the past we follow as closely as possible the method described by \cite{Barrows2016}, but instead of \textit{Chandra} X-ray observations we use a sample of compact radio sources.

\subsection{Detecting offset AGN: Radio or X-ray imaging?}

It is important for this work to have a reliable identification of the AGN. Low-luminosity AGN (LLAGN) can sometimes be confused with other sources such as X-ray binary systems (XRBs; stellar-mass black holes accreting mass from a companion star), ultraluminous X-ray sources (ULXs; either stellar mass black holes accreting at very high accretion rates [e.g. \citealt{King2001}], or intermediate-mass black holes [e.g. \citealt{Miller2004}]) or neutron stars. We argue that the distinction between AGN and other sources hosted by a galaxy can be made much more reliably using radio data than X-ray data, and for low-luminosity AGN the use of radio data is almost essential. Our argument is based upon the empirical relationship (known as the fundamental plane) between black hole mass, X-ray luminosity and radio luminosity \citep{Merloni2003} that applies across a wide range of black hole masses, from stellar-mass black holes to SMBHs, such that

\begin{equation}
{\rm log}L_{\rm R} = \left(0.60^{+0.11}_{-0.11}\right) {\rm log}L_{\rm X} + \left(0.78^{+0.11}_{-0.09}\right) {\rm log}M_{\rm BH}.
\end{equation}

Crucially, the term involving $M_{\rm BH}$ appears on the same side of the equation as $L_{\rm X}$, effectively boosting the radio luminosity of AGN, relative to their X-ray luminosities, by at least a few orders of magnitude. This boost from the $M_{\rm BH}$ term is considerably weaker for stellar- and intermediate-mass black holes, and AGN emission is therefore more likely to stand out in radio surveys than in X-ray surveys. Namely, if one assumes that the brightest source in the nucleus is likely to be the AGN, then this assumption is more likely to hold true at radio wavelengths than X-ray wavelengths.

X-ray searches for offset AGN may address the problem of confusion by imposing a minimum threshold on X-ray luminosity and/or hardness ratio, but in doing so most low-luminosity AGN would be excluded.

Additionally, radio imaging from an appropriate choice of interferometer, such as the Karl G. Jansky Very Large Array (VLA), in a suitable configuration, and at a high frequency, provides much higher angular resolution than X-ray imaging, enabling the identification of offsets that would be undetectable even by \textit{Chandra}.

\subsection{The accuracy of SDSS astrometry}

\cite{Orosz2013} matched 1\,297 compact objects from the International Celestial Reference Frame (ICRF2) catalog of radio sources, with positions localised using Very Long Baseline Interferometry (VLBI), with AGN positions from the SDSS data release 9 (DR9). They found the SDSS DR9 positions to be accurate to $\rm\simnot55~mas$ in both right ascension and declination, consistent with the declared SDSS astrometric precision.

The \cite{Orosz2013} histogram of positional differences takes the form of a Rayleigh distribution, but an extended tail to the distribution contains a significantly higher number of sources (51, or $\rm\simnot4~per~cent$ of the sample) with a $>3\sigma$ positional offset ($\rm>170~mas$) than would be expected statistically. The authors list other studies that have found similar fractions of radio/optical positional offsets (see \citealt{Orosz2013}, and references therein).

\subsection{The Cosmic-Lens All-Sky Survey (CLASS)}

Compared to X-ray AGN, radio AGN are less likely to be confused with emission from other types of objects hosted within a galaxy. However, a potential problem with selecting radio AGN is that, in additional to the nuclear radio emission, there can also be emission from jets and lobes. To eliminate this possible source of confusion, high-resolution radio observations are required to isolate the compact AGN core emission.

The CLASS survey \citep{Myers2003,Browne2003} was a successful hunt for radio-loud gravitationally lensed systems, that produced VLA observations of 13\,783 compact, flat-spectrum radio sources (16\,682 when combined with the VLA observations from the Jodrell Bank VLA Astrometric Survey [JVAS; \citealt{Patnaik1992}]) with declinations above 0 degrees. In order to ensure that emission from the majority of sources mapped were dominated by a single core component, each source targeted was selected to have a flat radio spectrum. The CLASS images were captured at 8.46 GHz with 200 to 250~milliarcsec (mas) angular resolution, and follow-up observations of likely lens candidates was performed using e-MERLIN and the Very Long Baseline Array (VLBA). The CLASS survey facilitated the discovery of 22 new gravitationally lensed systems.

From the 16\,682 images generated by CLASS, a total of 23\,418 individual components were identified. As expected, in the vast majority of cases the emission is dominated by a single, flat-spectrum, compact component which is the radio core. The additional components found are generally much weaker and of lower brightness than the dominant component, and most likely represent jet emission. The data, consisting of 8.46~GHz flux densities and their positions measured to an accuracy better than 0.1~arcsec are available online\footnote[2]{http://www.jb.man.ac.uk/research/gravlens/class/class.html}. Although originally a search for gravitationally lensed systems, the CLASS survey also serves as a convenient data repository for identifying offset AGN candidates.

In this current work we constructed our sample from the $z < 0.2$ OSSY catalog \citep[][]{Oh2011}, and cross matched the galaxy positions with the source positions in the CLASS survey. In basing our sample upon the OSSY catalog we are following the method used by \cite{Barrows2016} and \cite{Comerford2014} in their surveys. The OSSY catalog is a database of absorption- and emission-line measurements based upon the SDSS 7th data release of galaxies \citep{Abazajian2009}. We emphasize that we use the OSSY catalog in order to be able to compare our results with those of \cite{Barrows2016}, and also to use the OSSY emission-line data in order to reject starburst galaxies.

Throughout this paper we assume $H_0~=~{\rm 70~km~s^{-1}~Mpc^{-1}}$.

\section{The sample}
\label{sec-sample}

Our sample was constructed from the SDSS galaxy catalog by cross matching sources with the $z < 0.2$ OSSY catalog, and the CLASS catalog of compact radio sources. A list of sources cross-matched between the OSSY database and the SDSS galaxy catalog is available online\footnote[3]{http://gem.yonsei.ac.kr/$\sim$ksoh/wordpress/}, and this list was compared with entries in the CLASS database to find sources that lie within 10~arcsec. The cross-matched data were further reduced by visually inspecting the SDSS images, and eliminating any matches for which the CLASS detection was found to lie outside the visible bulge of the galaxy.

\begin{figure}
        \centering
	\includegraphics[width=85mm]{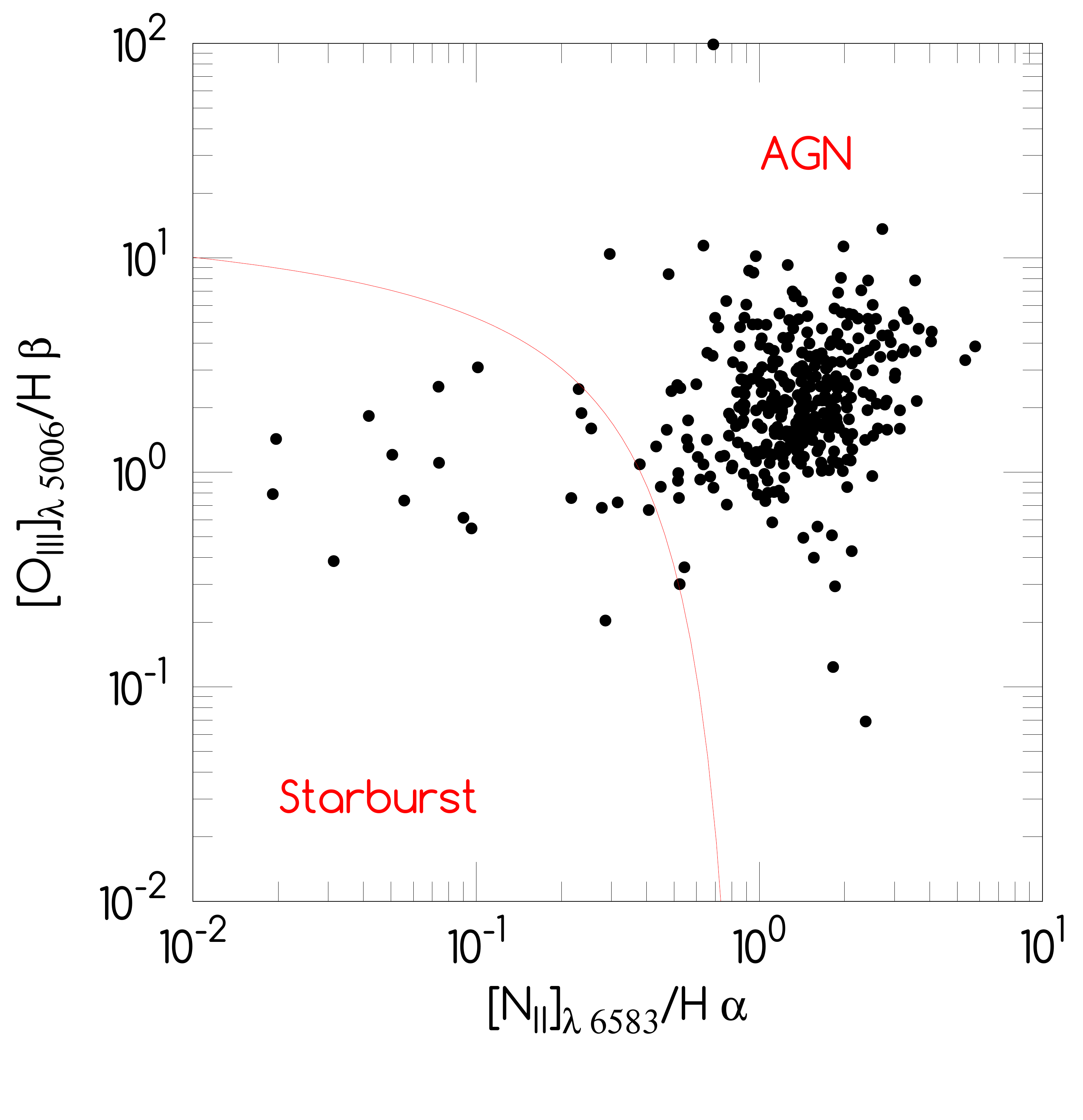}
        \caption{BPT diagram \protect\citep[see][]{Baldwin1981} for the 374 galaxies matched to CLASS positions, showing [N\,\textsc{ii}]$_{\lambda 6583}$/H$\alpha$ against [O\,\textsc{iii}]$_{\lambda 5006}$/H$\beta$. The solid red line is the AGN/starburst diagnostic curve suggested by \protect\cite{Kauffmann2003}. In total, 20 sources that fall within the starburst region of the plot were rejected from the sample.}
        \label{fig-bpt-diagram}
\end{figure}

Following this process our sample comprised 374 galaxies. To separate starburst galaxies from AGN galaxies we used the optical emission line strengths from the OSSY database to plot a BPT diagram \citep{Baldwin1981} of the [N\,\textsc{ii}]$_{\lambda 6583}$/H$\alpha$ ratio against the [O\,\textsc{iii}]$_{\lambda 5006}$/H$\beta$ ratio (see Fig.~\ref{fig-bpt-diagram}), which can be divided into separate starburst-dominated and AGN-dominated regions using an accepted demarcation curve. Our criteria for classifying galaxies as being starburst-dominated, and therefore rejecting them from our sample, was (from \citealt{Kauffmann2003}) 

\begin{equation}
\rm log([O\,\textsc{iii}]/H\beta) < [0.61 / (log([N\,\textsc{ii}]/H\alpha) - 0.05)] + 1.3,
\end{equation}
and
\begin{equation}
\rm log([N\,\textsc{ii}]/H\alpha) < 0.05.
\end{equation}

We subsequently rejected 20 galaxies  which we found to fall within the starburst region of the BPT diagram (although, strangely, 11 of these sources appear in the list of type-1 AGN published as part of the OSSY database), and a further 7 galaxies because one or more of the optical emission line data products were missing from the OSSY database, and classification was not possible. The vast majority of our sample - 347 out of 374 galaxies - were classified as AGN, which is unsurprising given that these galaxies are all found to contain a compact radio source in their nucleus.

We additionally rejected two further galaxies, SDSS~J095551.73+694048.6 (M82) and SDSS~J112017.01+133522.8 (NGC~3628), which, despite being classified as AGN in our BPT diagnostic plot, are well-known starburst galaxies. Despite the CLASS survey being limited to compact sources with flat spectra, a few extended sources crept in as a result of the difficulty in getting reliable radio spectral indices for extended sources. For reasons of transparency, we include these galaxies in some of the later discussions. The final sample therefore numbers 345 galaxies.

Let us assume that the uncertainties in RA and dec are normally distributed for both the SDSS and CLASS positions, and have variances $\sigma_{\rm sdss}^2$ and $\sigma_{\rm class}^2$ respectively. We would then expect the difference in position to follow a statistical distribution that we could predict, provided there are no AGN offsets.

The distribution of the difference between two normally distributed values is itself normally distributed, and our measured offsets in RA and dec (which we denote $\Delta_{\rm R}$ and $\Delta_{\rm D}$ respectively) should therefore have Gaussian distributions (denoted $P_{\Delta_{\rm R}}$ and $P_{\Delta_{\rm D}}$) with variance $\sigma^2 = \sigma_{\rm sdss}^2 + \sigma_{\rm class}^2$. 

The distribution of the radial offset, $\Delta = \sqrt{\Delta_{\rm R}^2 + \Delta_{\rm D}^2}$, is denoted as $P_{\Delta}$, and given by

\begin{equation}
P_{\Delta} = \int\limits_{-\infty}^{\infty} \int\limits_{-\infty}^{\infty} P_{\Delta_{\rm R}} \, P_{\Delta_{\rm D}} \, \delta \left( \Delta - \sqrt{\Delta_{\rm R}^2 + \Delta_{\rm D}^2} \right) d\Delta_{\rm R} \, d\Delta_{\rm D},
\end{equation}

which evaluates to a Rayleigh distribution with variance $\sigma^2 = \sigma_{\rm sdss}^2 + \sigma_{\rm class}^2$,

\begin{equation}
P_{\Delta} = \frac{\Delta}{\sigma^2} \, {\rm exp} \left( -\frac{\Delta^2}{2 \sigma^2} \right).
\end{equation}

\begin{figure}
        \centering
	\includegraphics[width=85mm]{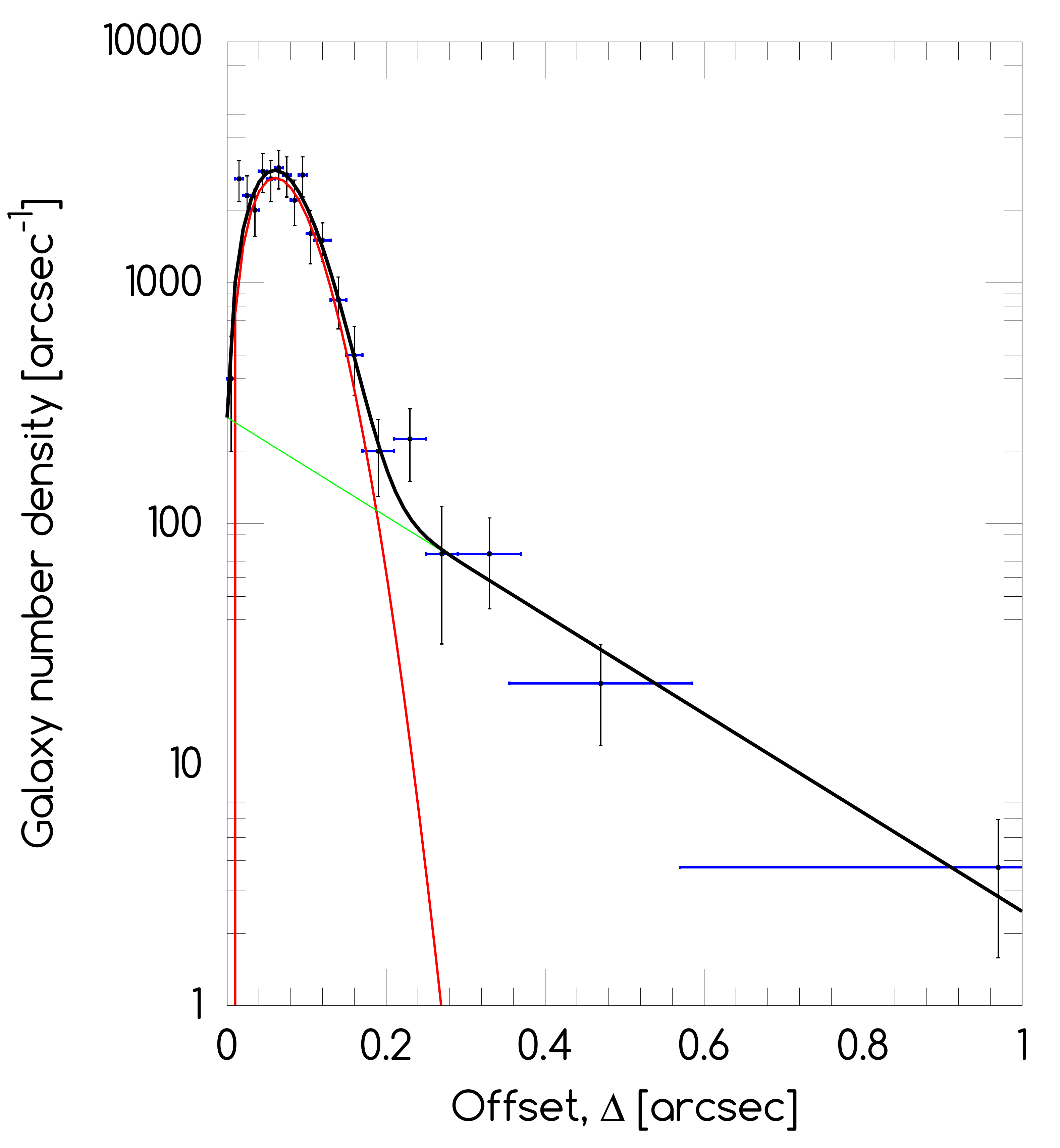}
        \caption{Histogram of offsets between the SDSS and CLASS positions for the 345 galaxies in our sample. The data have been fitted with a Rayleigh distribution with $\sigma = 60.5~{\rm mas}$ (thick red line) and an exponential component of the form $\rm exp(-4.72 \Delta)$ (narrow green line). The sum of these two components is shown with a thick black line. The x-axis error bars show the widths of the bins used, and the fit quality was $\chi^2_{\rm red} = 0.81$, for 20 degrees of freedom.}
        \label{fig-offset-histogram}
\end{figure}

A histogram of the measured radial offsets is shown in Figure \ref{fig-offset-histogram}, and the data were originally fitted with a Rayleigh distribution (see also \citealt{Makarov2017, Orosz2013}) with $\sigma = 59.2~{\rm mas}$. The fit was reasonably good ($\chi^2_{\rm red} = 1.88$ for 22 degrees of freedom), but at offsets greater than 110~mas there was a slight excess in the number of galaxies, above the number expected from the Rayleigh distribution fit, in most bins. We therefore fitted the data with a model consisting of two components:

\begin{enumerate}

\item A Rayleigh distribution (with the normalisation and $\sigma$ as free parameters; the best fit for the latter was $\sigma = {\rm 60.5~mas}$) in order to model the offsets that result from the uncertainty in the SDSS and CLASS positions (see Fig.~\ref{fig-offset-histogram}, thick red line).

\item An exponential component, of the form $\rm exp (\lambda \Delta)$, to model any offsets that are not described by the Rayleigh distribution (see Fig.~\ref{fig-offset-histogram}, thin green line). We offer no astrophysical justification for the choice of an exponential function, except that it provides a remarkably good fit to the data. The best fit to the decay constant was $\lambda = \rm-4.72~arcsec^{-1}$. This component describes the offset distribution of sources that may include offset AGN (and also chance alignments with background AGN, and other radio detections that are not the result of offset AGN), and serves as a useful upper limit on the predicted number of offset AGN in our sample.

\end{enumerate}

The fit quality of the two-component model is very good, with $\chi^2_{\rm red} = 0.81$ for 20 degrees of freedom. For a given offset, the probability that this offset is the result of positional uncertainty can be estimated using the ratio of the Rayleigh distribution to the sum of both components, and, throughout this paper, all such probability estimates have been derived using this method.

\section{Results}

The CLASS survey obtained its source positions using the VLA in "A" configuration, with an angular resolution of 200 to 250~mas at 8.46~GHz \citep{Myers2003}, while astrometric precision for compact components was better than 100~mas. Meanwhile, the galaxy positions listed in the SDSS galaxy catalog were derived from fitting the galaxy profiles with suitable models (using different filters), and their accuracy depends upon the regularity of the galaxy shapes, and upon the angular size of a galaxy. Large galaxies will usually have larger uncertainties on their centroid positions. We therefore expect that some of the observed offsets are simply the result of positional uncertainty rather than a spatially offset SMBH.

\begin{figure}
        \centering
	\includegraphics[width=85mm]{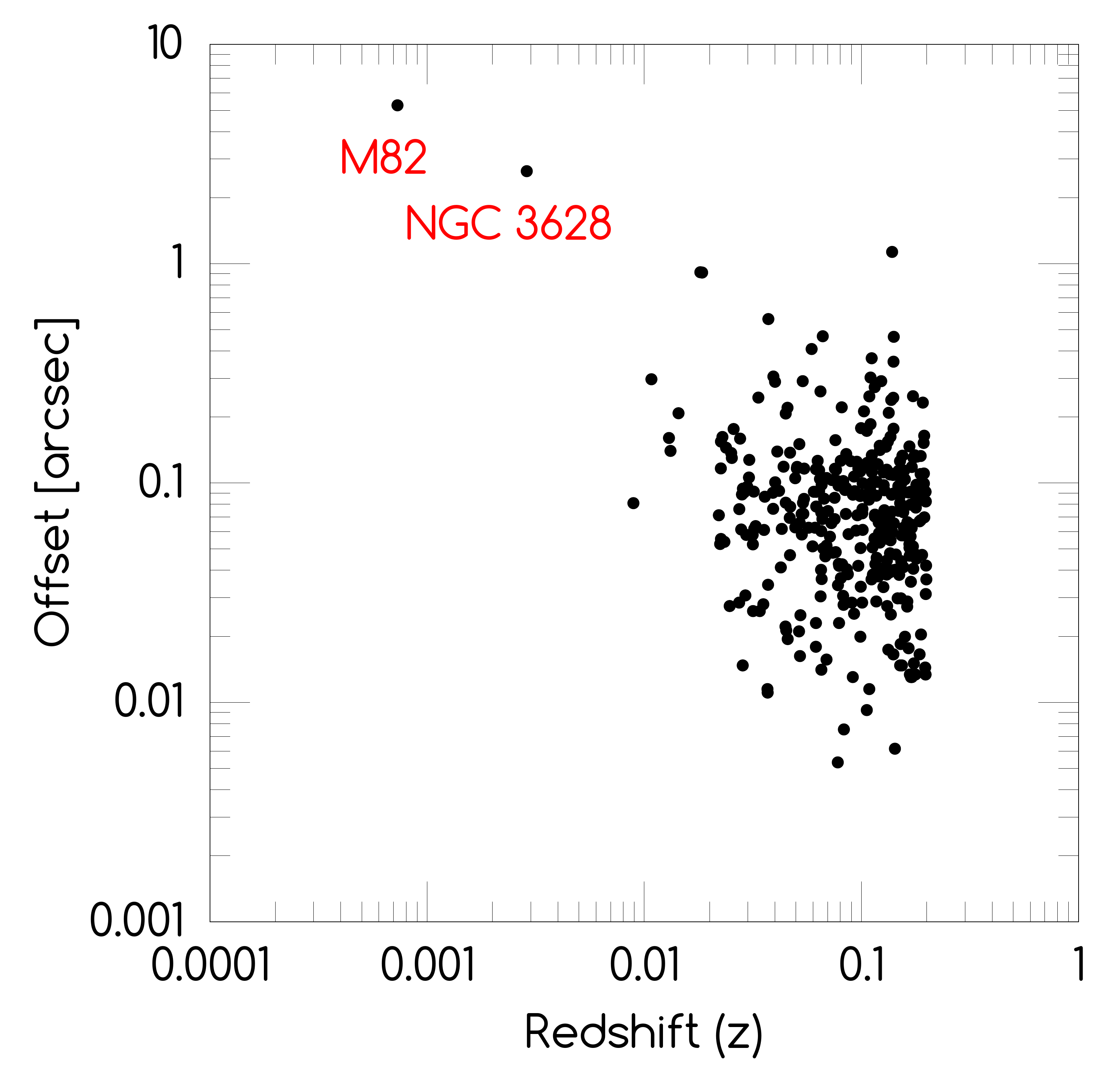}
        \caption{Redshift (from the SDSS galaxy catalog) versus CLASS/SDSS offset for our sample of 345 galaxies. We also include M82 and NGC~3628, which we have rejected from the sample on the basis that they are well-known, nearby starburst galaxies. The remaining galaxies show only a weak anti-correlation between redshift and offset (Pearson correlation coefficient $r = -0.127$).}
        \label{fig-redshift-v-offset}
\end{figure}

Since nearby galaxies will, based upon an equivalent spatial offset, exhibit larger angular offsets than more distant galaxies, we plot a scatter diagram of redshift against angular offset (Fig.~\ref{fig-redshift-v-offset}). Two of the larger offsets are in well-known nearby starburst galaxies (M82 [3.5~Mpc; \cite{Dalcanton2009}] and NGC~3628 [11.9~Mpc; \cite{Shapley2001}]), and although these galaxies may both host low-luminosity AGN \citep{Matsumoto1999,Dahlem1995} we choose to exclude them from our statistics, reducing the total number of galaxies in our sample to 345. The remaining galaxies show a weak anti-correlation between redshift and offset (Pearson correlation coefficient $r = -0.127$). The mean redshift for the galaxies with radio offsets of at least 150~mas is 0.087, which is very similar to the mean redshift of 0.082 found for the 18 galaxies in the sample of \cite{Barrows2016}.

In Section~\ref{sec-sample} we fitted the histogram of SDSS/CLASS offsets with a Rayleigh function, representing the statistical distribution of offsets due to positional uncertainty, and an exponential tail with decay constant $\lambda = \rm-4.72~arcsec^{-1}$, representing the remainder of the sample. The fitted uncertainty for the Rayleigh component was $\sigma = {\rm 60.5~mas}$, and assuming an SDSS positional uncertainty of 55~mas \citep{Orosz2013}, and ignoring any other potential sources of error, implies that the uncertainty in the CLASS positions is $\rm\simnot25~mas$.

By integrating the exponential function we place an upper limit of 59 on the number of offset AGN in our sample (17 per cent). Around half of these galaxies are found at offsets of at least 150~mas, where the number of detections (44) is almost 250 per cent greater than predicted by the Rayleigh distribution (13). At offsets of less than 150~mas our fit is dominated by the Rayleigh component, and we therefore choose to focus our attention on the 38 offset-AGN candidates with offsets greater than 150~mas, and flux densities of at least 8~mJy (listed in Table~\ref{tbl-sample-with-offsets}, and shown in the appendix). For these sources, which we hereafter refer to as the high-offset galaxies, we predict that positional uncertainty accounts for fewer than 30 per cent of the offsets, so our chances of observing offset AGN are good.

The mean offset was found to be 99~mas, which corresponds to 0.27 of the mean VLA beam size in X-band (A-configuration), and only 0.071 of the mean SDSS point-spread function (PSF), where the mean beam size for the 38 high-offset galaxies was found to be 362~mas in length and 280~mas in width, and the SDSS PSF in r-band is 1.4~arcsec.

\begin{table*}
	\caption{The high-offset galaxies: galaxies identified as AGN, and with the closest matched CLASS source offset by an angle of at least 150~mas, and with a flux density of at least 8~mJy. The CLASS flux is shown only for the brightest source detected in the galaxy, except when there are two or more sources with similar fluxes. An indication is shown where there is a fainter source with a smaller offset. The full table, including all 345 galaxies in the sample, is published online in the electronic edition of this journal.}
	\label{tbl-sample-with-offsets}
	\begin{center}
	\begin{tabular}{@{}lcccccccc}
		\hline
		SDSS ID & OSSY ID & z$^{a}$ & CLASS ID & \multicolumn{2}{c}{CLASS/SDSS\,$^{b}$} & Flux\,$^{b,c}$ & Alternative Name \\
		& & & & \multicolumn{2}{c}{Offset} & (8.46 GHz) \\
		& & & & [mas] & [pc]$\,^{d}$ & [mJy] \\
		\hline
		J075244.19+455657.3 & 587737808493936696$\,^{e}$ & 0.052 & GB6J075241+455641 & 150 & 158 & 75.4 \\
		J075614.47+272435.6 & 588013382720422440 & 0.140 & GB6J075619+272437 & 245 & 663 & 14.4 \\
		J082905.46+140406.7 & 587741816772493607 & 0.137 & GB6J082906+140350 & 239 & 635 & 22.3 \\
		J083824.01+254516.4 & 587738947736633359 & 0.019 & GB6J083823+254526 & 911 & 346 & 9.6 & NGC~2623; Arp 243 \\
		J084002.36+294902.5 & 587735240637284507 & 0.065 & GB6J084000+294859 & 262 & 341 & 10.1 \\
		J084307.11+453742.8 & 587729387137925339 & 0.192 & GB6J084306+453739 & 232$\,^{f}$ & 839 & 135.0 \\
		J091445.53+413714.2 & 588013382729859273 & 0.141 & GB6J091445+413658 & 357 & 971 & 9.8 \\
		J093346.08+100908.9 & 587735344799350868 & 0.011 & GB6J093346+100924 & 297 & 66 & 24.1 & NGC~2911; Arp 232 \\
		J093551.58+612111.7 & 587725551741370430$\,^{e}$ & 0.039 & GB6J093551+612102 & 306 & 245 & 31.2 & UGC~5101 \\
		J102544.22+102230.4 & 587734948056727714 & 0.046 & GB6J102545+102246 & 220 & 204 & 50.0 \\
		J103258.89+564453.3 & 587729386609967112 & 0.045 & GB6J103303+564334 & 207 & 189 & 28.8 & MCG~+10-15-099 \\
		J111125.21+265748.9 & 587741602562310202 & 0.034 & GB6J111125+265738 & 245 & 168 & 20.3 & NGC~3563B \\
		J111916.53+623925.7 & 588009372835709000 & 0.110 & GB6J111912+623938 & 303 & 654 & 45.5 \\
		J112039.94+504938.2 & 588013382203211796 & 0.028 & GB6J112040+504930 & 159 & 90 & 13.8 \\
		J115000.07+552821.3 & 587731870170349772 & 0.139 & GB6J114959+552832 & 1\,131 & 3\,031 & 70 \\
		J115133.06+050605.2 & 588010879295029355 & 0.076 & GB6J115133+050609 & 157 & 238 & 27.6 \\
		J115410.41+122509.7 & 588017703466762347 & 0.081 & GB6J115409+122522 & 221 & 358 & 187.5 \\
		J115905.67+582035.5 & 587735697516331091 & 0.054 & GB6J115859+582034 & 291 & 316 & 9.5 & MCG~+10-17-121 \\
		J122209.28+581421.5 & 587731891653181635 & 0.100 & GB6J122208+581427 & 178 & 350 & 35.0 \\
		J122513.09+321401.5 & 587739608627937400 & 0.059 & GB6J122513+321406 & 408 & 485 & 80.5 \\
		J122622.47+640622.0 & 587729153598881886 & 0.110 & GB6J122621+640637 & 185 & 401 & 70.7 \\
		J124135.08+285036.5 & 587741531719270424 & 0.066 & GB6J124136+285035 & 467 & 623 & 31.4 \\
		J125433.25+185602.1 & 588023668102004902 & 0.115 & GB6J125433+185557 & 273 & 617 & 128.2 \\
		J131503.51+243707.7 & 587741726584143934 & 0.013 & GB6J131459+243611 & 160 & 43 & 10.3 & IC~860 \\
		J133435.06+344639.9 & 587739305297444932 & 0.026 & GB6J133437+344630 & 176 & 93 & 20.9 & NGC~5228; UGC~8556 \\
		J133621.18+031951.0 & 587726033869668455 & 0.023 & GB6J133617+031909 & 162 & 76 & 21.5 & MCG~+01-35-014 \\
		J134243.62+050432.1 & 587729159508066374 & 0.136 & GB6J134242+050430 & 163 & 429 & 168.7 \\
		J134442.15+555313.8 & 587735666377949228$\,^{e}$ & 0.037 & GB6J134444+555322 & 559 & 425 & 33.6 & Mrk~273; UGC~8696 \\
		J135022.12+094010.7 & 587736543088672829 & 0.132 & GB6J135023+094011 & 154 & 397 & 250.4 \\
		J135036.01+334217.3 & 587739304762081386 & 0.014 & GB6J135037+334228 & 208 & 62 & 95.8 & NGC~5318; UGC~8751 \\
		J135927.60+465045.9 & 587735429621022820 & 0.173 & GB6J135926+465047 & 249 & 817 & 10.2 \\
		J142730.26+540923.7 & 587735697525375126 & 0.106 & GB6J142724+540920 & 173 & 360 & 19.5 \\
		J144441.09+142346.9 & 587742628524327082 & 0.141 & GB6J144444+142347 & 177 & 480 & 22.9 \\
		J144607.04+090338.2 & 587736543631573219 & 0.194 & GB6J144607+090349 & 152 & 554 & 76.2 \\
		J153457.21+233013.3 & 587739815850606701 & 0.018 & GB6J153456+233006 & 1\,687$\,^{g}$ & 629$\,^{g}$ & 60.1$\,^{g}$ & IC~4553; Arp 220 \\
		J154912.33+304716.4 & 587736919430922502 & 0.112 & GB6J154912+304712 & 2\,103$\,^{g}$ & 4\,603$\,^{g}$ & 38.0$\,^{g}$ \\
		J162719.15+483126.7 & 587725993572434085 & 0.195 & GB6J162718+483134 & 164 & 601 & 32.5 \\
		J215527.22+120500.6 & 587727221937078784 & 0.109 & GB6J215528+120444 & 248 & 531 & 58.8 \\
		\hline
	\end{tabular} \\
	\end{center}
	\textsc{Notes}: $^{a}$ Redshift, from the OSSY catalog, $^{b}$ Source shown is the brightest CLASS source matched to this galaxy; an indication is shown if fainter sources with smaller offsets are found, $^{c}$ From the CLASS survey \citep{Myers2003,Browne2003}, $^{d}$ Based upon distances estimated from the redshift (OSSY catalog), $^{e}$ Galaxy appears in the OSSY catalog of type-1 AGN, $^{f}$ \cite{Orosz2013} find a slightly lower offset of 177~mas between the International Celestial Reference Frame (ICRF2) radio source and the SDSS position, $^{g}$ Other, fainter, sources have smaller offsets (NB: All CLASS detections in the galaxies shown here have offsets that are at least 150~mas). \\
\end{table*}

We refer the reader to the appendix for comments on individual sources, but a few sources are worth highlighting:

\begin{itemize}

\item Four of the strongest candidates for hosting offset AGN are SDSS~J153457.21+233013.3 (IC~4553, Arp~220), SDSS~J083824.01+254516.4 (NGC~2623, Arp~243), SDSS~J134442.15+555313.8 (Mrk~273) and SDSS~J093551.58+612111.7 (UGC~5101). All are well-known merging galaxies \citep[e.g.][]{Xia2002, Joseph1985, Casoli1988, Joy1987, Armus2004, Lonsdale2003}, and it has been previously suggested \citep{U2013} that Mrk~273 hosts a dual AGN system.

\item We find a few less-well-known galaxies that appear to be excellent candidates for offset-AGN hosts: SDSS~J115000.07+552821.3 (1.13~arcsec offset), SDSS~J124135.08+285036.5 (467~mas offset), SDSS~J082905.46+140406.7 (239~mas offset), SDSS~J115410.41+122509.7 (221~mas offset), and SDSS~J122513.09+321401.5 (408~mas offset). The offset of the radio source in SDSS~J115000.07+552821.3 has also been noted by \cite{DeVries2009}.

\item The lack of any \textit{Chandra} detection of SDSS~J154912.33+304716.4 and SDSS~J131503.51+243707.7 (IC~860) is contrary to what we would expect from an AGN host galaxy, as even a Compton-thick AGN can usually be expected to have some reprocessed emission in the soft X-ray band. \cite{Lehmer2010} established an upper limit of only ${\rm log} L_{\rm 2-10\,keV} = {\rm 40.19~erg~s^{-1}}$ on the X-ray luminosity of IC~860 (\textit{Chandra} observation 10\,400, 19.1~ks exposure).

\end{itemize}

Overall, the 38 galaxies selected have interesting and diverse properties. At least five are clearly interacting galaxies, four have dust lanes, several are disk galaxies - which is very unusual for the host of a radio AGN - and one turns out to be a radio lobe lensed by an intervening galaxy.

\subsection{Probability of chance alignments}

To derive a rough estimate of the probability of a chance alignment between a background radio source and an unrelated foreground galaxy we use the total number of OSSY galaxies with declinations greater than zero degrees, which is $\simnot610\,000$, and multiply this number by a search area around each galaxy, for which we choose to use a circle of radius 2~arcsec. When we measure the angular size of the galaxy bulges in the SDSS images from the appendix (r-band) we find an average radius of 1.4~arcsec, so our choice of 2~arcsec is generous.

We therefore estimate that the OSSY galaxies with declinations greater than zero degrees occupy $\rm\simnot0.6~deg^{2}$, or $\rm\simnot3 \times 10^{-3}~per~cent$ of the northern sky. The probability of a radio source being unrelated to the galaxy with which it has been matched is around 1 in 35\,000, and there is roughly a 1.2~per~cent chance that at least one of the 433 CLASS sources we have matched to our sample of 345 galaxies is actually a background source.

\section{Discussion}

\subsection{Overview}

We have used the CLASS archive and the SDSS galaxy catalog to identify 345 AGN-hosting galaxies with compact, flat-spectrum radio sources in their cores. We believe that there is good evidence for some radio AGN being offset from the centroid of the optical emission, as revealed by SDSS. It is important to emphasize that this does not necessarily imply a true offset AGN; i.e. that the radio source does not lie at the centre of mass of its host galaxy. The offset-AGN candidates presented in this paper include a number of interesting objects, some of which are already known to be unusual, or suspected of hosting offset AGN.

By integrating the non-Rayleigh component of our offset histogram we place a strong upper limit on the fraction of offset AGN of around 17 per cent, which represents the fraction of our sample for which the offsets cannot be explained by statistical error in the measurements of the optical centroid and radio position. In reality the percentage is likely to be much smaller than this because in our reduced sample of 38 galaxies we have 11 galaxies with an obviously disturbed morphology, and four more galaxies with visible dust lanes, both of which mean the optical centroids are unlikely to be a good measure of the dynamical centres of these galaxies. We therefore estimate a slightly less-strong upper limit of around 10 per cent.

\subsection{Comparison with \protect \cite{Barrows2016} X-ray survey}

We can compare our results with those obtained using other search methods. Firstly, we find that only three of the 345 galaxies (0.87 per cent) in our sample (SDSS~J115000.07+552821.3, SDSS~J153457.21+233013.3 [Arp 220] and SDSS~J083824.01+254516.4 [NGC~2623]) have offsets of more than 600~mas, compared to 18 out of a sample of 48 (37.5 per cent) reported by \cite{Barrows2016} from their \textit{Chandra} survey.

Expressing the offsets in terms of spatial separation, rather than angular separation, also suggests some differences between the two surveys: \cite{Barrows2016} report that the spatial offsets they detect are estimated to be between 0.8 and 19.4~kpc, whereas we find only six galaxies in our sample with offsets in this range, which represents less than two per cent of the total sample size.

In rough order of astronomical interest we list several possible reasons for the discrepancy between X-ray and radio searches for spatially-offset AGN:

\begin{itemize}

\item A significant fraction of the measured offsets represent real offset AGN. In this case a possible explanation for the difference in the frequency of real offsets in X-ray- and radio-selected samples could be attributed to X-ray loudness happening early in the merger process, while radio loudness is a late-time-phenomenon. Radio and X-ray surveys may therefore be sampling different evolutionary stages of the merger process. Delayed triggering of radio AGN has already been invoked in a different context by \cite{Shabala2017}.

\item Very few of the measured offsets represent real offset AGN. In this case one explanation might be that the host galaxies of X-ray-loud AGN are more optically disturbed than the host of radio-loud AGN, and thus optical centroids of these galaxies are unreliable. Examination of the SDSS images of the \cite{Barrows2016} candidates suggests that many lie in merging, or at least disturbed, galaxies. Again, as above, this fits with an evolutionary scenario.

\item Some of the X-ray sources claimed to be AGN are ULXs or objects unrelated to the AGN emission.

\item X-ray surveys find higher offset-AGN fractions because they generally exclude low-luminosity AGN (\citealt{Barrows2016}, for example, require that the 2-10~keV luminosity exceeds $\rm 10^{42}~erg~s^{-1}$). \cite{Comerford2014} found that the fraction of offset AGN candidates increases with bolometric luminosity (in the bolometric range $10^{43}$ to $\rm 10^{46}~erg~s^{-1}$).

\item Radio astrometry is much better than X-ray astrometry, and some of the X-ray claims arise as a result of underestimating the X-ray position errors.

\end{itemize}

The reality is that it is probable that a combination of these possibilities is required to account for the results, and the idea that the discrepancy between X-ray and radio surveys has an astrophysical origin should be treated with scepticism.

\subsection{Comparison with \protect \cite{Comerford2014} survey of velocity offsets}

However, the upper limit of 17 per cent that we place on offset-AGN candidates in our sample is more than double that reported by \cite{Comerford2014}, who found 351 offset-AGN candidates from a sample of 18\,314 Type-2 AGN by looking for velocity offsets in optical spectra, and estimate that once projection effects in observed velocities are corrected then between 4 and 8 per cent of AGN are offset.

Of the 345 galaxies in our sample only four are also found in the \cite{Comerford2014} list of offset-AGN candidates, and in two of those galaxies - SDSS~J091445.53+413714.2 and SDSS~J124135.08+285036.5 - we find large offsets of 357 and 467~mas respectively (see the appendix for SDSS images). These offsets deviate from the fitted Rayleigh distribution with 5-sigma confidence, so we regard these galaxies as especially high-priority follow-up targets.

\subsection{Conclusion/Summary}

We summarise our results as follows:

\begin{itemize}

\item By integrating the non-Rayleigh component of our offset histogram we place an upper limit on the fraction of radio AGN that are offset from the optical centroids of their host galaxies of 17 per cent, but expect that many of these candidates are the result of poorly-constrained optical centroids due to disturbed or irregular host morphologies.

\item We find that only three from our sample of 345 galaxies (0.87~per~cent) have radio offsets of more than 600~mas, compared to 18 out of a sample of 48 (37.5~per~cent) reported by \cite{Barrows2016} from their \textit{Chandra} survey.

\item We have identified a number of interesting new offset-AGN candidates, and, in addition, our sample contains a diverse collection of sources, including disturbed/interacting galaxies, spiral galaxies, and galaxies with dust lanes.

\end{itemize}

Optical/radio offsets are not sufficient, on their own, to identify offset AGN, and therefore our data cannot distinguish between sources with optical/radio offsets caused by measurement error, and sources with offsets that may be genuine. However, such offsets are very helpful in identifying suitable candidates for follow-up observations, in which better optical measurements can help to more accurately locate the galaxy centroid. In a follow-up paper we shall examine in more detail some of the better offset-AGN candidates from our sample.

\section{Acknowledgements}

This research has made use of the NASA/IPAC Extragalactic Database (NED), which is operated by the Jet Propulsion Laboratory, California Institute of Technology, under contract with the National Aeronautics and Space Administration. The National Radio Astronomy Observatory is a facility of the National Science Foundation operated under cooperative agreement by Associated Universities, Inc.

\bibliographystyle{mnras}
\bibliography{references}

\appendix

\clearpage
\section{Summary of individual sources}
\label{sec-individual sources}

In this appendix we show SDSS images, X-band VLA images (all 'A' configuration), and, where available, \textit{Chandra} images, of the 38 galaxies with CLASS detections of at least 8~mJy in flux density, offset from the SDSS positions by an angle of at least 150~mas. Assuming calibration has been done correctly then the uncertainty in the position of a VLA source scales roughly with the beam size, and inversely with the signal-to-noise ratio (see \citealt{White1997} for a more detailed discussion of positional accuracy). We expect an 8~mJy source to have an uncertainty of around 20~mas in its position, and we attempt to limit our CLASS positional uncertainties to no greater than 20~mas by excluding six galaxies from this appendix that contain only sources with flux densities lower than 8~mJy.

The green squares in the full-colour SDSS images highlight the regions that are shown in more detail in the red-filter SDSS and \textit{Chandra} images. The green crosses mark the SDSS galaxy positions, and the blue circles (which use different shades of blue for visibility purposes) show the positions of the CLASS sources. All the blue circles have a radius of 500~mas. The contours in the VLA images are at 1~mJy, 10~mJy and 100~mJy flux densities.

All spectra referred to in this appendix were obtained from the SDSS Data Release 12. To identify Seyfert galaxies we use the demarcation curve of \cite{Kewley2006}, such that a galaxy is classified as a Seyfert if

\begin{equation}
\rm log([O\,\textsc{iii}]/H\beta) > (1.18 \times log([O\,\textsc{i}]/H\alpha)) + 1.3,
\end{equation}

where the $\rm [O\,\textsc{i}]_{\lambda 6300}$ and $\rm [O\,\textsc{iii}]_{\lambda 5006}$ lines from the OSSY database were used. We find eight Seyfert galaxies amongst these 38 high-offset sources. The remaining galaxies are either unclassified due to missing emission-line data, or are classified as low-ionisation emission-line regions (LINERs). However, we hesitate to label these objects as LINERs because the SDSS spectra often resemble those of passive-elliptical galaxies, with little evidence of LINER activity, and many of the emission lines described in the OSSY database are undetectable.

\subsection{SDSS~J075244.19+455657.3}

\begin{flushleft}
Morphology: E (NED)
\end{flushleft}
Brightest and nearest CLASS source 75.4~mJy (6C B074906.2+460422), offset by 150~mas from the SDSS position. Compact \textit{Chandra} source detected at the same position. The spectrum has a broad component to H$\alpha$, and this source appears in the OSSY catalog of type 1 AGN.

\vspace{0.2cm}

\noindent\begin{minipage}{0.48\textwidth}
	\centering
	\includegraphics[width=56mm]{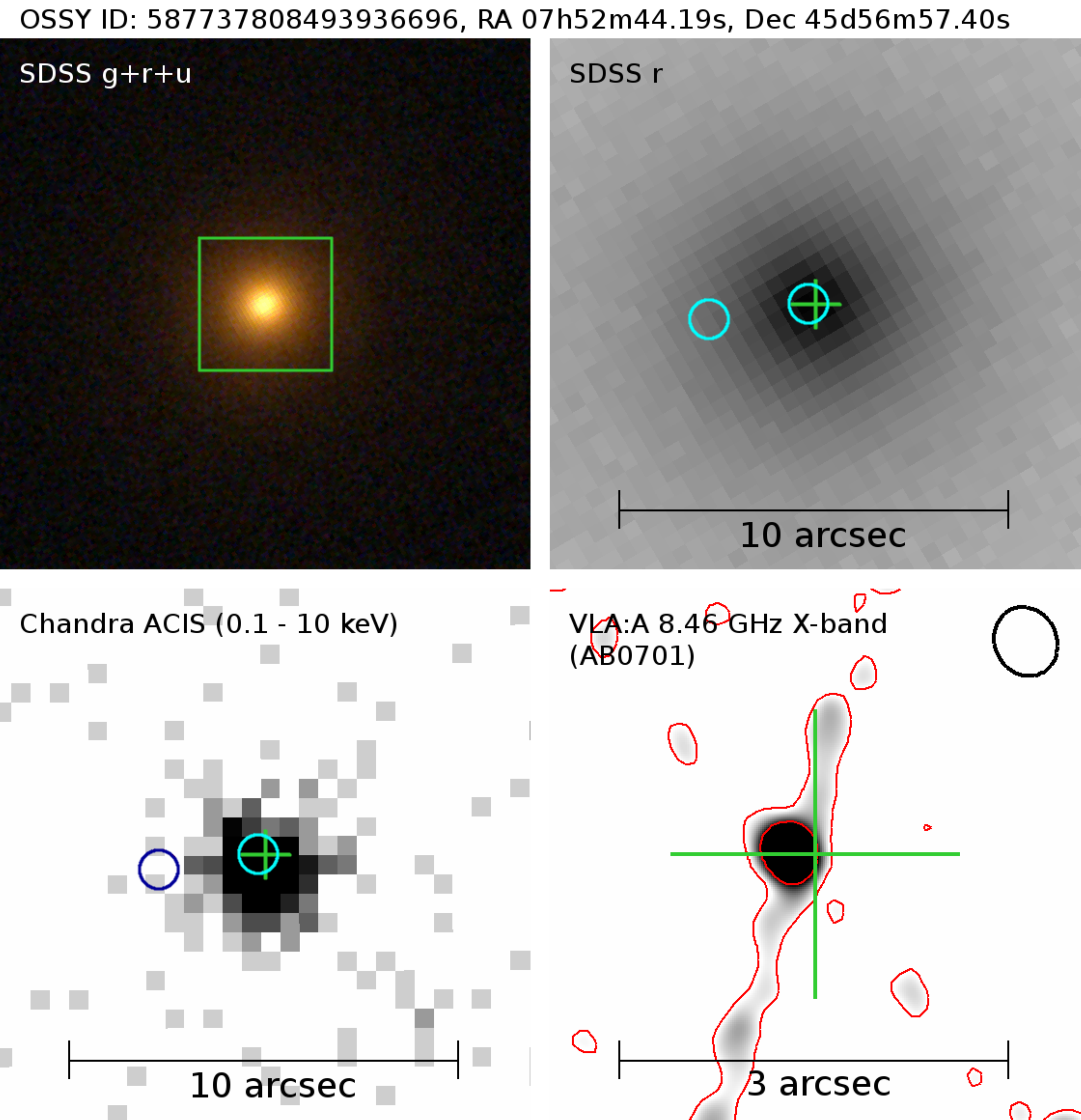}
	\begin{flushright}
	Chandra obs: 18226
	\end{flushright}
\end{minipage}

\subsection{SDSS~J075614.47+272435.6}

Brightest and nearest CLASS source of 14.4~mJy (NVSS~J075614+272436), offset by 245~mas from the SDSS position. Two fainter sources of up to 0.6~mJy also detected further from the nucleus.

\vspace{0.2cm}

\noindent\begin{minipage}{0.48\textwidth}
	\centering
	\includegraphics[width=85mm]{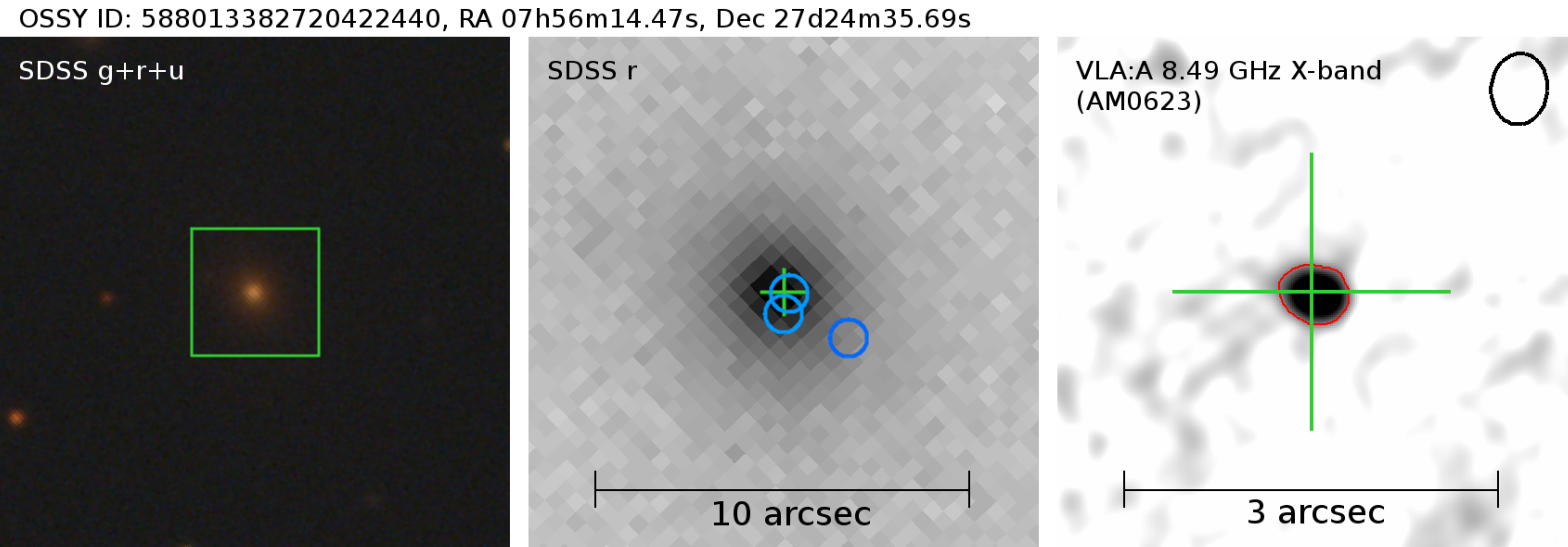}
\end{minipage}

\subsection{SDSS~J082905.46+140406.7}

Single CLASS detection of 22.3~mJy (NVSS~J082905+140407), offset by 239~mas from the SDSS position. The SDSS spectrum looks like that of a passive galaxy, but the OSSY line strengths indicate that this galaxy may be a LINER. A potential spiral galaxy, and a strong candidate for an offset-AGN host.

\vspace{0.2cm}

\noindent\begin{minipage}{0.48\textwidth}
	\centering
	\includegraphics[width=85mm]{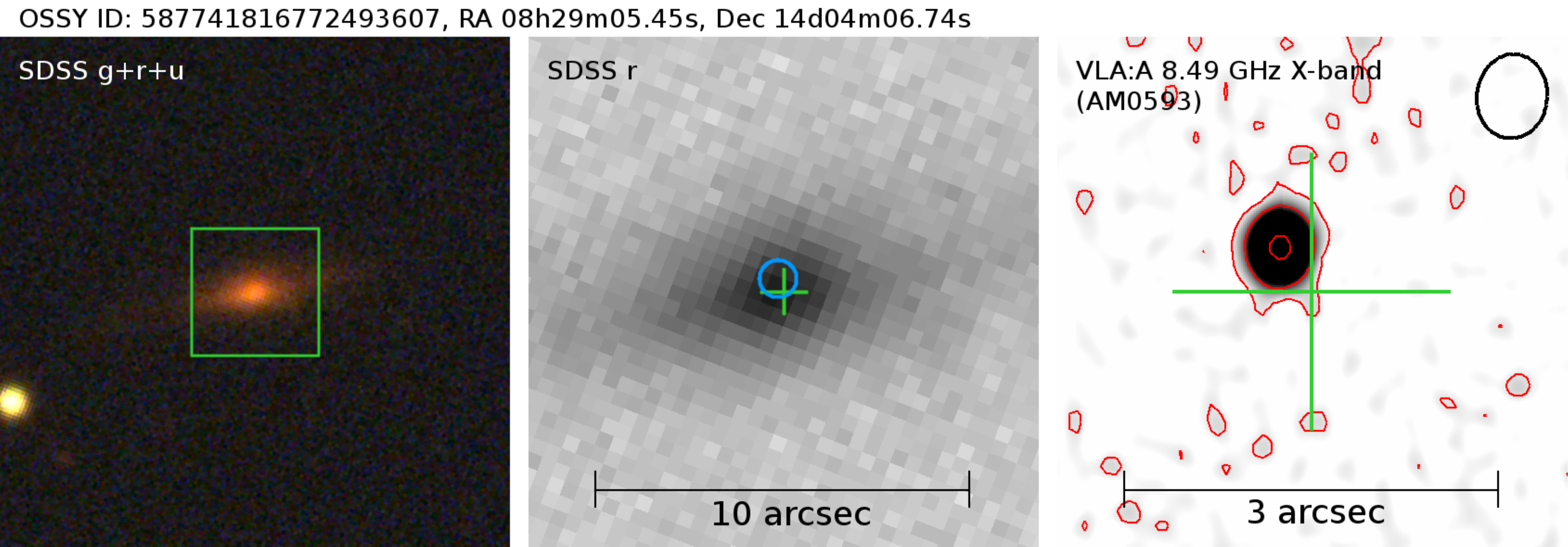}
\end{minipage}

\subsection{SDSS~J083824.01+254516.4 (NGC~2623; Arp 243)}

\begin{flushleft}
Morphology: pec (NED)
\end{flushleft}
There is a cluster of three CLASS detections with similar luminosities: 8~mJy (1.20~arcsec offset), 9.6~mJy (0.91~arcsec offset) and 13.6~mJy (1.21~arcsec offset), one or more of which could be related to NVSS~J083824+254516. A fainter source of 0.7~mJy lies further away. The \textit{Chandra} detection is strong, unresolved, and is offset from the optical centre and coincident with the CLASS detections. NGC~2623 is a well-known merging galaxy \citep{Arp1966, Joseph1985, Casoli1988, Joy1987}. These offsets deviate from the fitted Rayleigh distribution with 5-sigma confidence.

\vspace{0.2cm}

\noindent\begin{minipage}{0.48\textwidth}
	\centering
	\includegraphics[width=56mm]{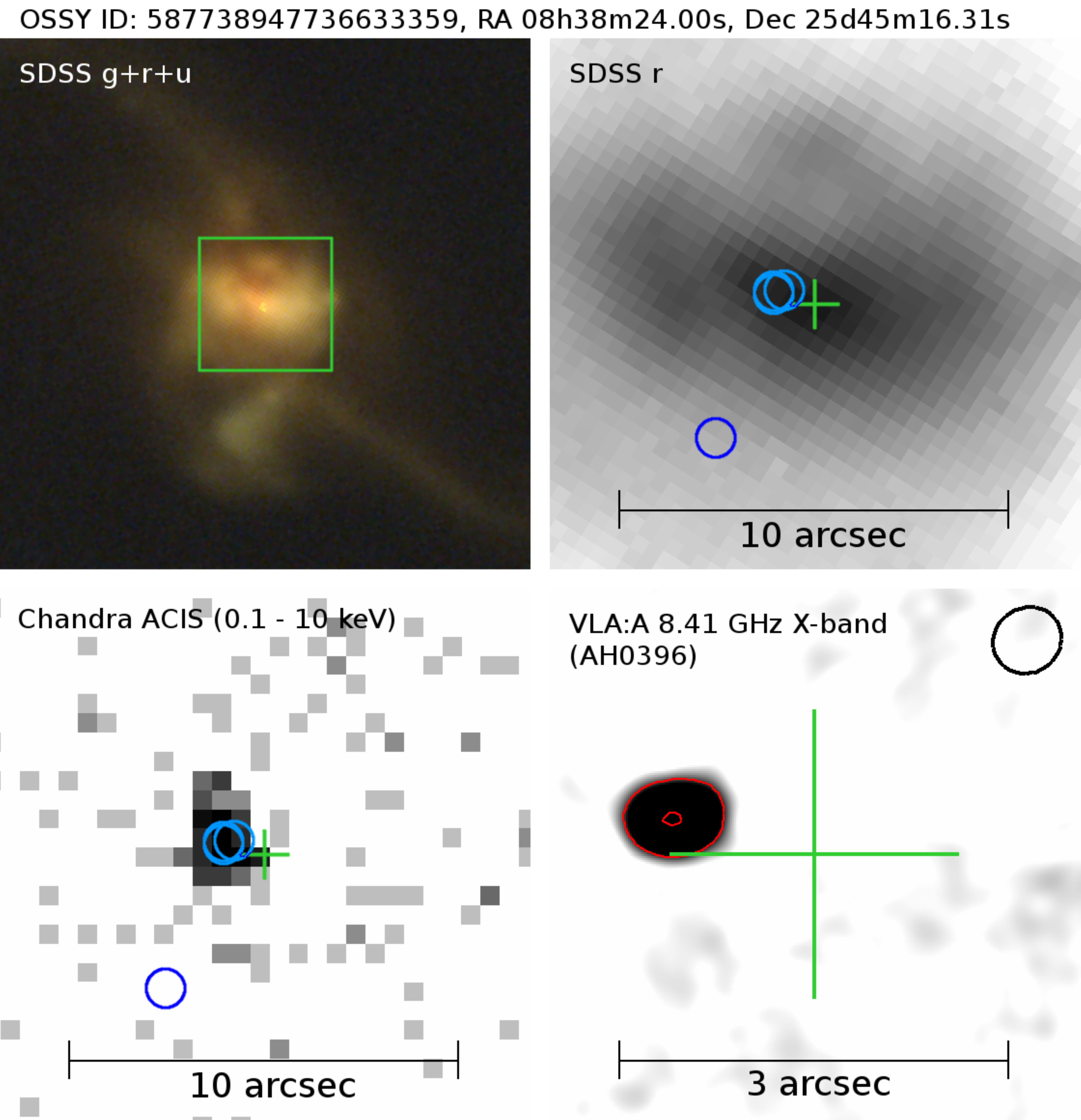}
	\begin{flushright}
	Chandra obs: 4059
	\end{flushright}
\end{minipage}

\subsection{SDSS~J084002.36+294902.5}

\begin{flushleft}
AGN class: Seyfert, Morphology: diffuse, Sa (NED)
\end{flushleft}
Single CLASS detection of 10.1~mJy (4C +29.30), offset by 262~mas from the SDSS position. The optical images show a red nucleus surrounded by a blue haze. A second object, offset by approximately 5~arcsec, is not detected by CLASS or \textit{Chandra}. The spectrum shows exceptionally strong [O\,\textsc{iii}] emission compared to other lines, and there is no visible {4000~\AA} break. The AGN is classified as a Seyfert~2 by \cite{Veron-Cetty2006}. The CLASS detection is the core of an extended radio galaxy.

\vspace{0.2cm}

\noindent\begin{minipage}{0.48\textwidth}
	\centering
	\includegraphics[width=56mm]{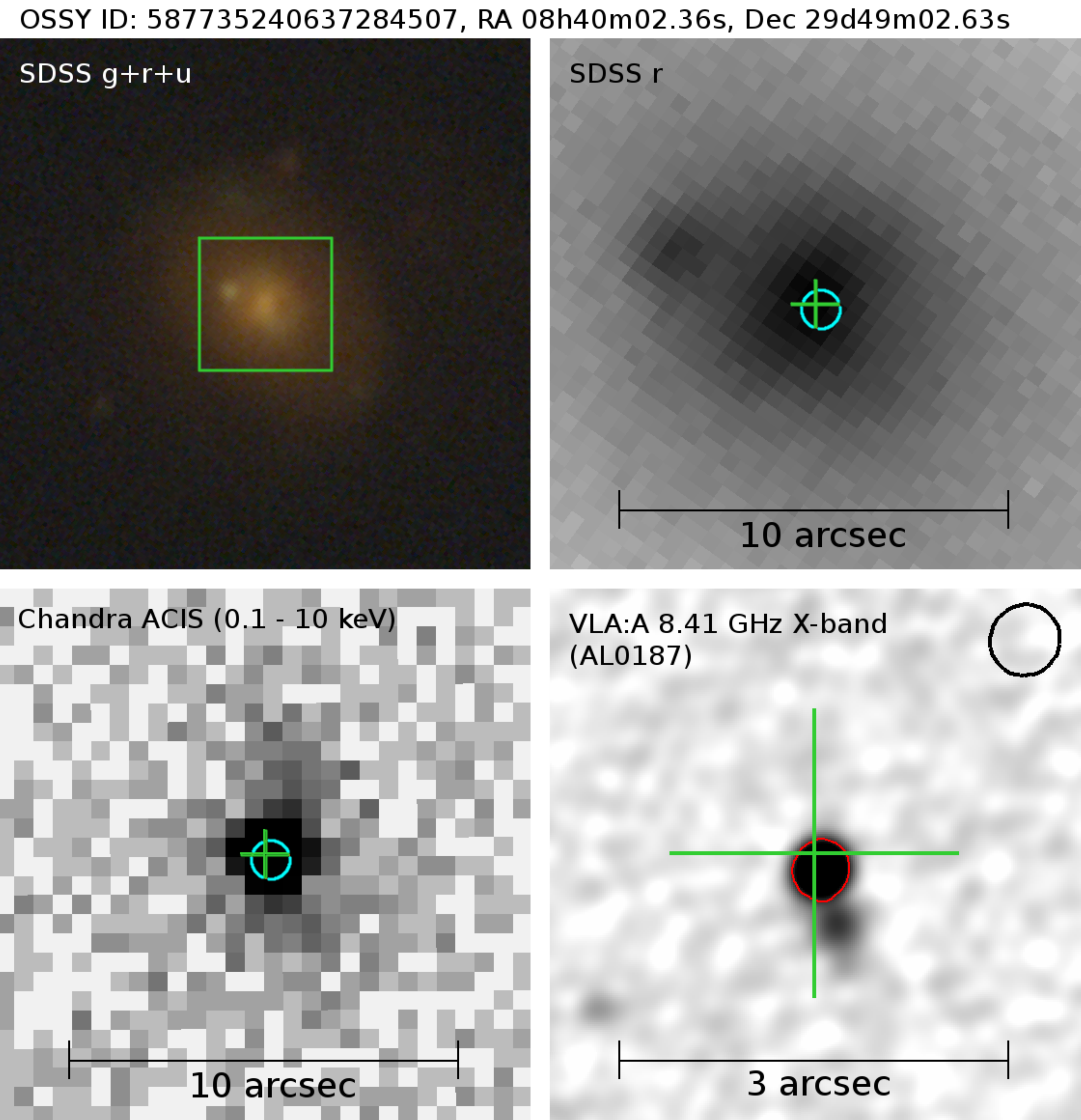}
	\begin{flushright}
	Chandra obs: 11688
	\end{flushright}
\end{minipage}

\subsection{SDSS~J084307.11+453742.8}

Single, bright, CLASS detection of 135~mJy (NVSS~J084307+453743), offset by 232~mas from the SDSS position. \cite{Orosz2013} find a slightly lower offset of 177~mas for this source. The SDSS images suggest that this galaxy may possibly host a disk. Although the SDSS spectrum has no obvious emission lines, the source was classified as an AGN using the OSSY data. However, OSSY failed to find the $\rm [O\,\textsc{i}]_{\lambda 6300}$ line, so the type of AGN is unknown. This galaxy is a promising candidate for an offset-AGN host, but requires high-resolution optical/infra-red observations to improve the centroid position.

\vspace{0.2cm}

\noindent\begin{minipage}{0.48\textwidth}
	\centering
	\includegraphics[width=85mm]{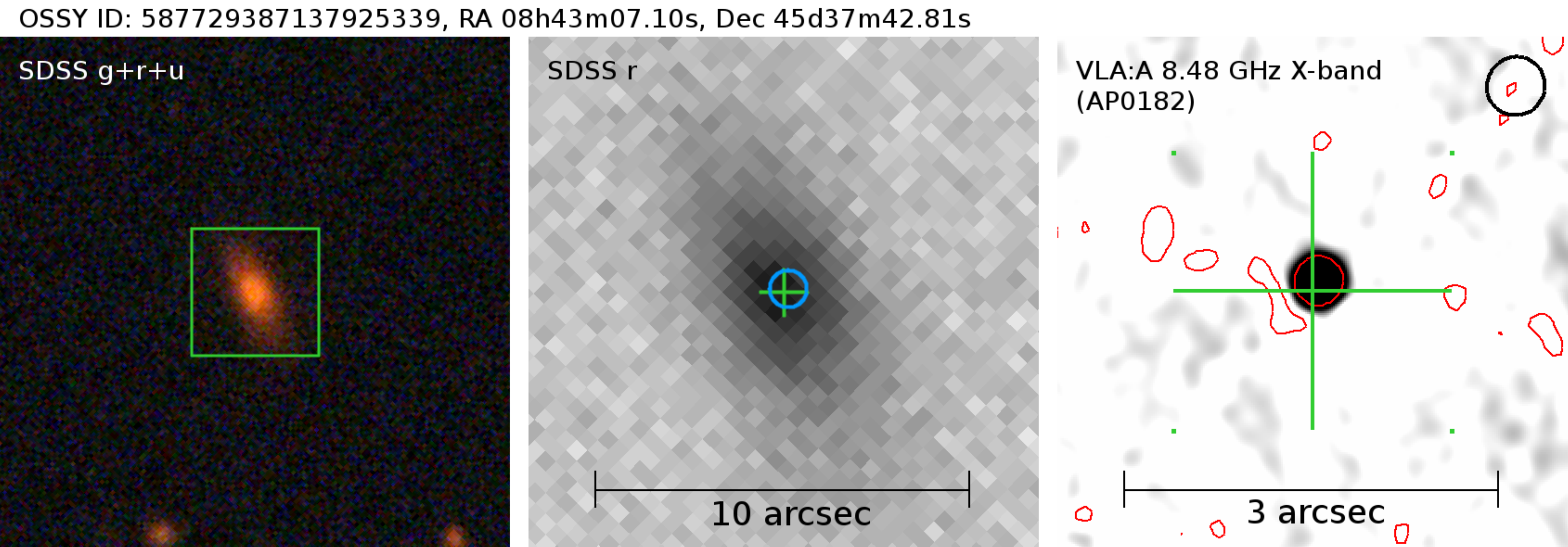}
\end{minipage}

\subsection{SDSS~J091445.53+413714.2}

Single CLASS detection of 9.8~mJy (B3 0911+418), offset by 357~mas from the SDSS position. This object was selected by \cite{Comerford2014} as an offset-AGN candidate on the basis of a spectral-line offset. There is a clear dust lane, which would make the optical centre more difficult to determine, but it still represents a good follow-up candidate. This offset deviates from the fitted Rayleigh distribution with 4-sigma confidence.

\vspace{0.2cm}

\noindent\begin{minipage}{0.48\textwidth}
	\centering
	\includegraphics[width=85mm]{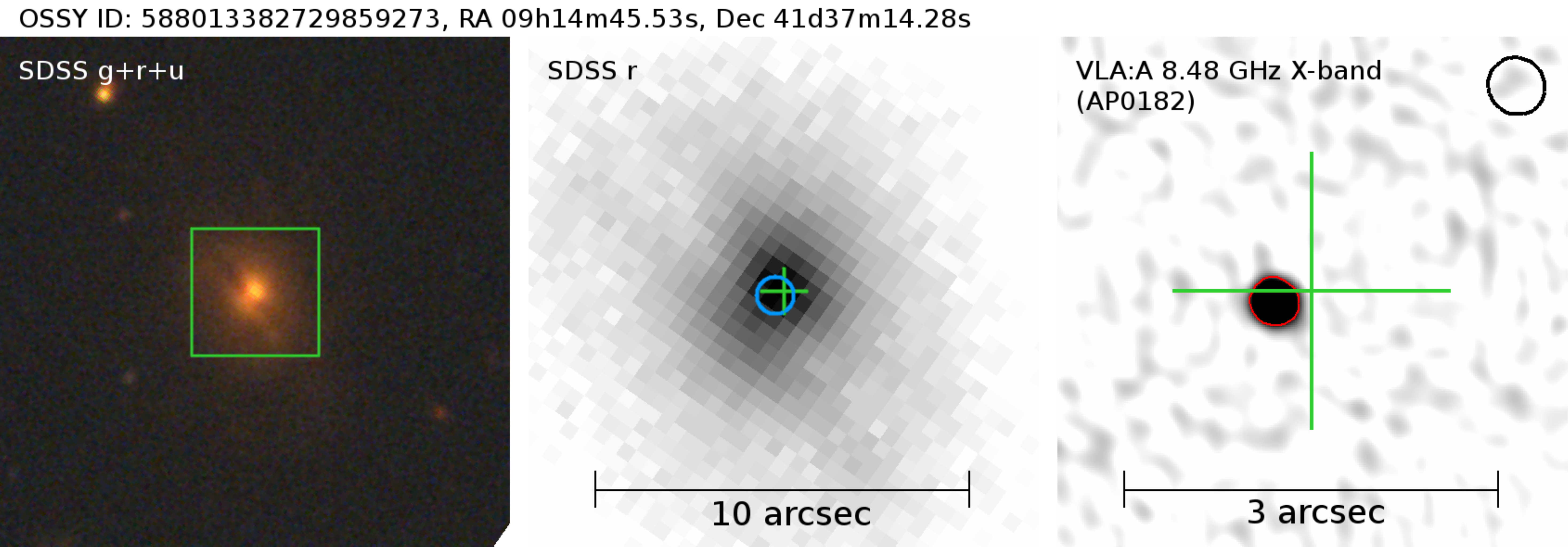}
\end{minipage}

\subsection{SDSS~J093346.08+100908.9 (NGC~2911; Arp 232)}

\begin{flushleft}
Morphology: SA0(s)? pec, S0 pec, E (NED)
\end{flushleft}
Single CLASS detection of 24.1~mJy (NVSS~J093346+100909), offset by 297~mas to the south-east of the SDSS position. This galaxy is at the centre of the NGC~2911 group, and a well-known LINER with an active radio core \citep{Filippenko1985,Filho2002,Schilizzi1983,Mezcua2014}. This offset deviates from the fitted Rayleigh distribution with 4-sigma confidence.

\vspace{0.2cm}

\noindent\begin{minipage}{0.48\textwidth}
	\centering
	\includegraphics[width=85mm]{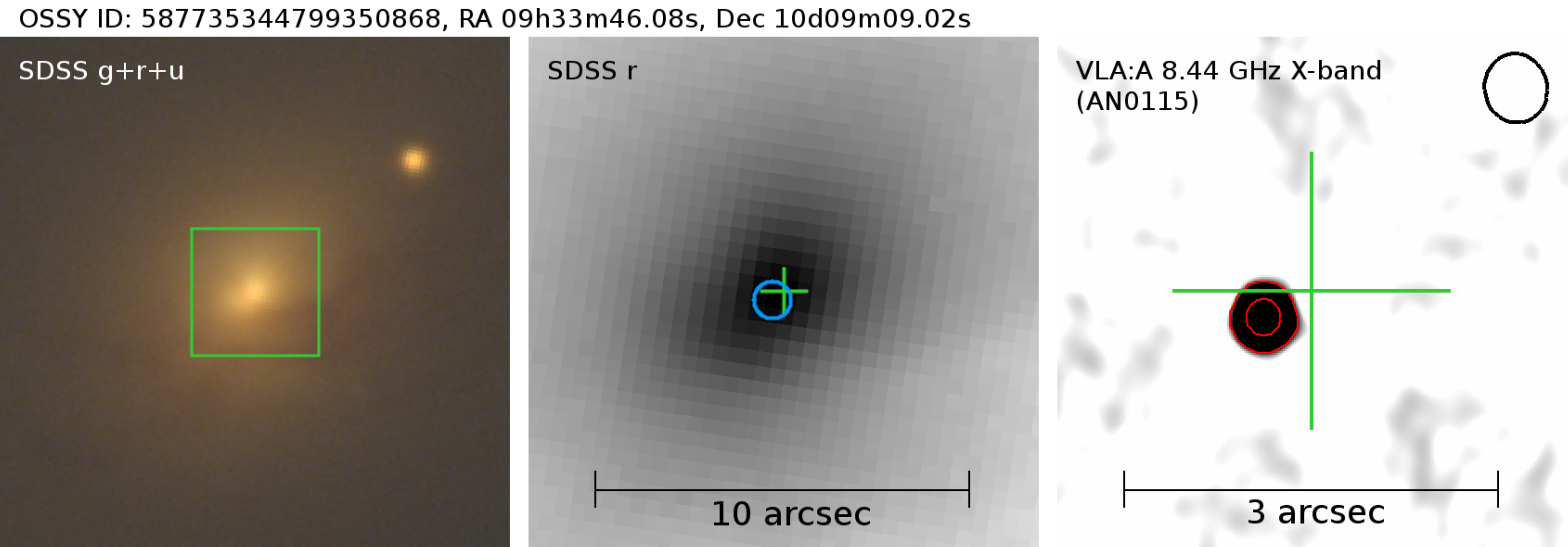}
\end{minipage}

\subsection{SDSS~J093551.58+612111.7 (UGC~5101; MCG~+10-14-025)}

\begin{flushleft}
AGN class: Seyfert, Morphology: S?, Sb, pec (NED)
\end{flushleft}
Two CLASS detections of 31.2~mJy and 17.4~mJy, with almost the same position, are offset by 306 and 339~mas respectively (one or both of which may be NVSS~J093551+612112) from the SDSS position. Compact \textit{Chandra} source detected at the same position. UGC~5101 is a well-known interacting galaxy \citep[e.g.][]{Armus2004, Lonsdale2003}, and the AGN is classified as a Seyfert~1 by \cite{Veron-Cetty2006}. These offsets deviate from the fitted Rayleigh distribution with 4-sigma confidence.

\vspace{0.2cm}

\noindent\begin{minipage}{0.48\textwidth}
	\centering
	\includegraphics[width=56mm]{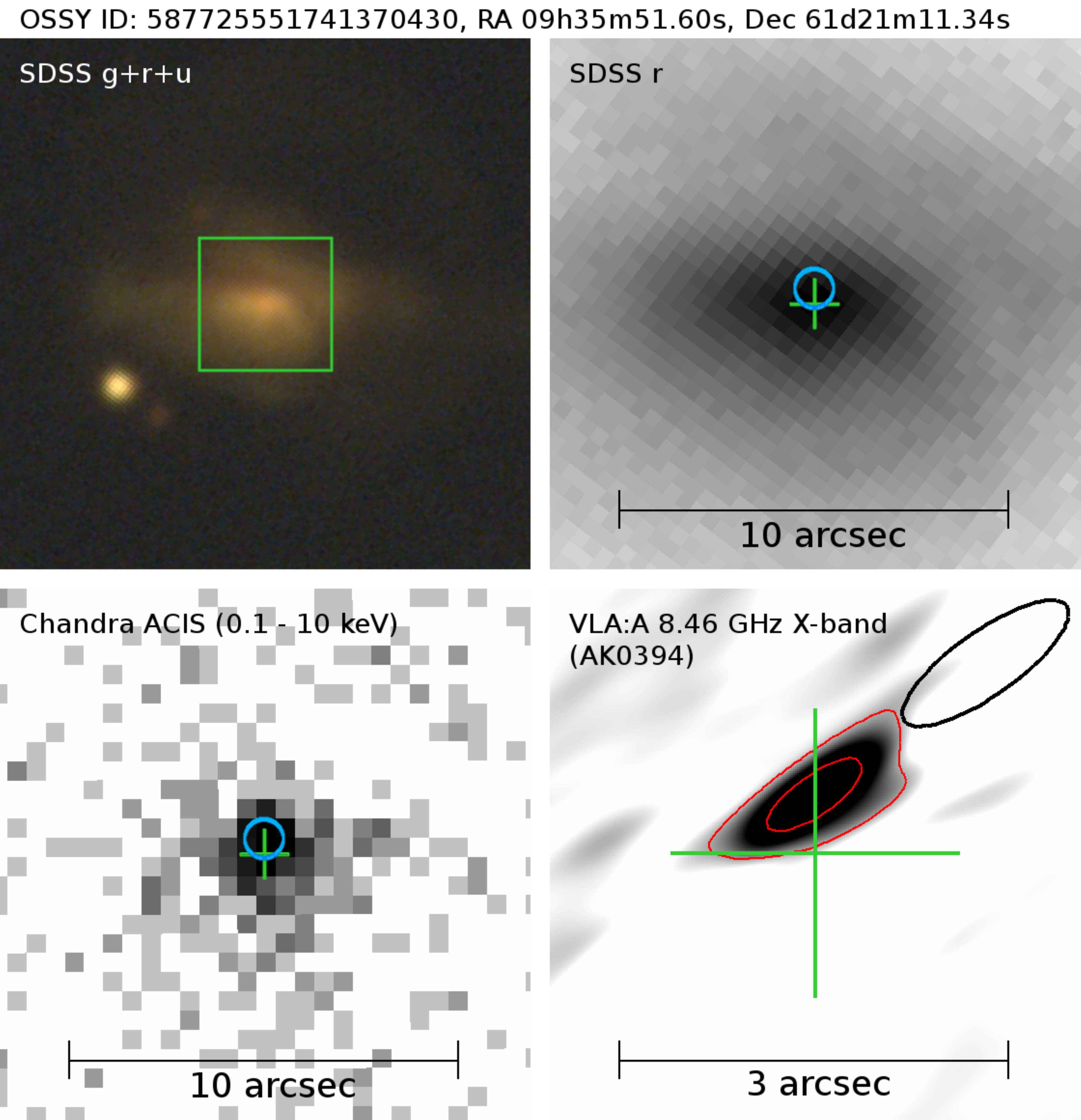}
	\begin{flushright}
	Chandra obs: 2033
	\end{flushright}
\end{minipage}

\subsection{SDSS~J102544.22+102230.4}

\begin{flushleft}
Morphology: S0 (NED)
\end{flushleft}
Single CLASS detection of 50.0~mJy (NVSS~J102544+102231), offset by 221~mas from the SDSS position. A prominent dust lane is likely to make the position of the optical centre more difficult to locate.

\vspace{0.2cm}

\noindent\begin{minipage}{0.48\textwidth}
	\centering
	\includegraphics[width=85mm]{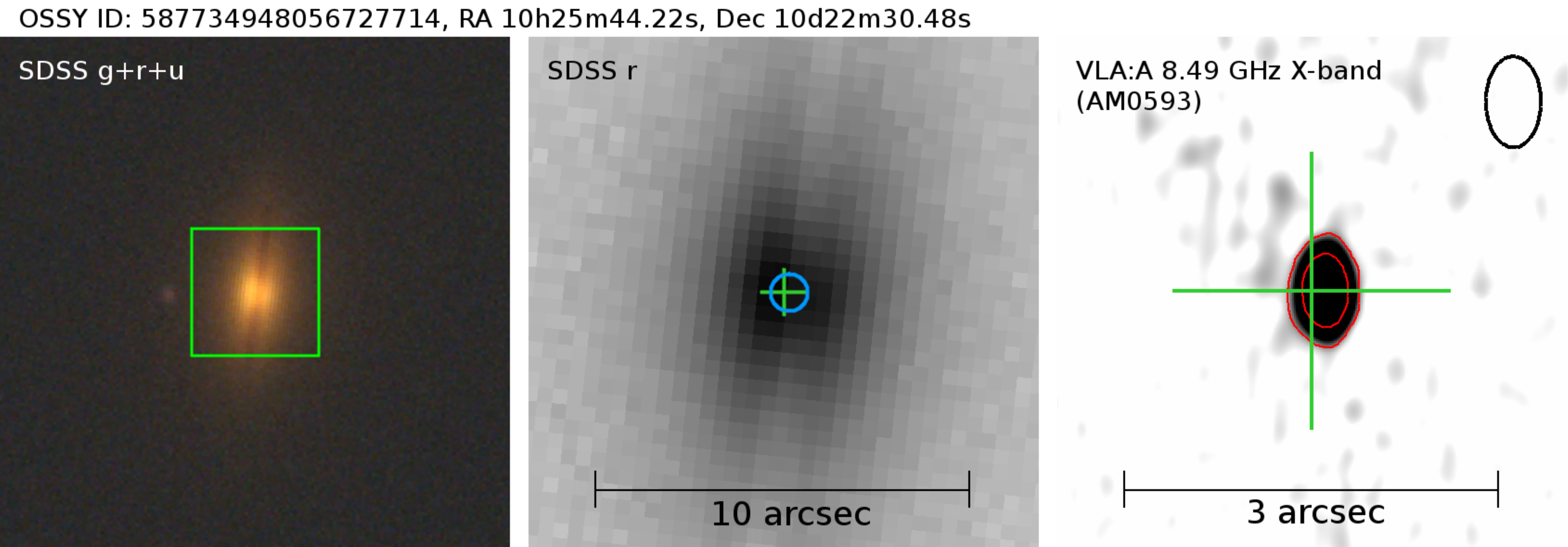}
\end{minipage}

\subsection{SDSS~J103258.89+564453.3 (MCG~+10-15-099)}

\begin{flushleft}
Morphology: compact (NED)
\end{flushleft}
Single CLASS detection of 28.8~mJy (7C 1029+5700), offset by 207~mas from the SDSS position. This galaxy has a very prominent core in a diffuse halo, and the SDSS spectrum appears to be that of a passive elliptical galaxy. However, OSSY line strengths indicate that this galaxy may be a LINER. We consider this galaxy a strong candidate for an offset-AGN host.

\vspace{0.2cm}

\noindent\begin{minipage}{0.48\textwidth}
	\centering
	\includegraphics[width=85mm]{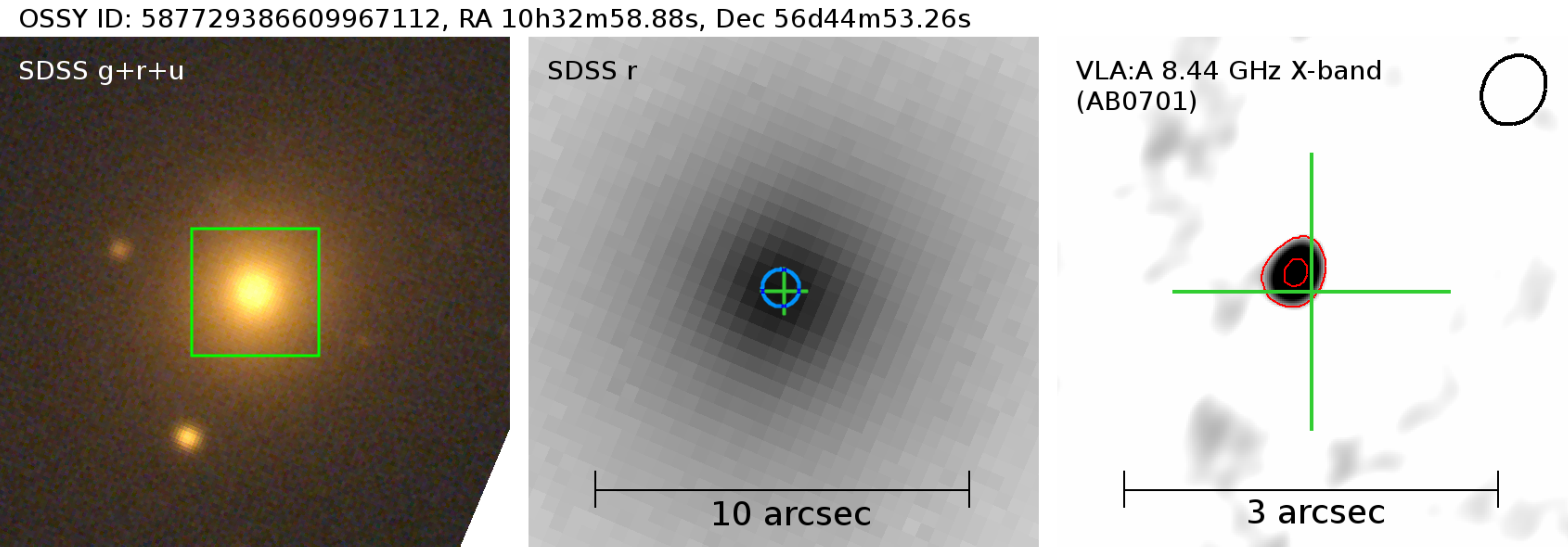}
\end{minipage}

\subsection{SDSS~J111125.21+265748.9 (NGC~3563B; MCG~+05-27-014)}

\begin{flushleft}
Morphology: SB0?, SB?0 (NED)
\end{flushleft}
Single CLASS detection of 20.3~mJy (CRATES~J1111+2657), offset by 245~mas from the SDSS position. Part of the NGC~3563 pair of interacting galaxies, and hosts a dust lane to the west of the nucleus which could make the true optical centre difficult to locate.

\vspace{0.2cm}

\noindent\begin{minipage}{0.48\textwidth}
	\centering
	\includegraphics[width=85mm]{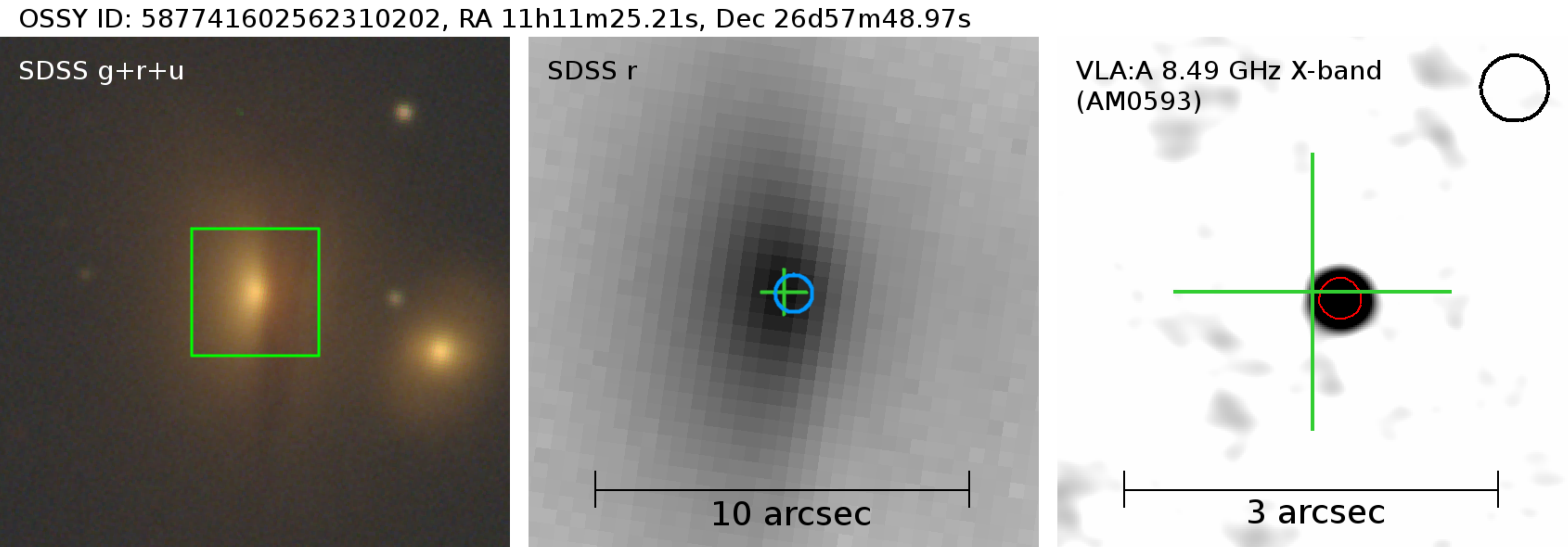}
\end{minipage}

\subsection{SDSS~J111916.53+623925.7}

Single CLASS detection of 45.5~mJy (NVSS~J111916+623926), offset by 303~mas from the SDSS position. The SDSS spectrum is that of a passive elliptical with little evidence of AGN activity, but the OSSY line strengths indicate that this galaxy may be a LINER. Optical images hint at a morphology of possibly S0 or E7, with a bright nucleus. This offset deviates from the fitted Rayleigh distribution with 4-sigma confidence.

\vspace{0.2cm}

\noindent\begin{minipage}{0.48\textwidth}
	\centering
	\includegraphics[width=85mm]{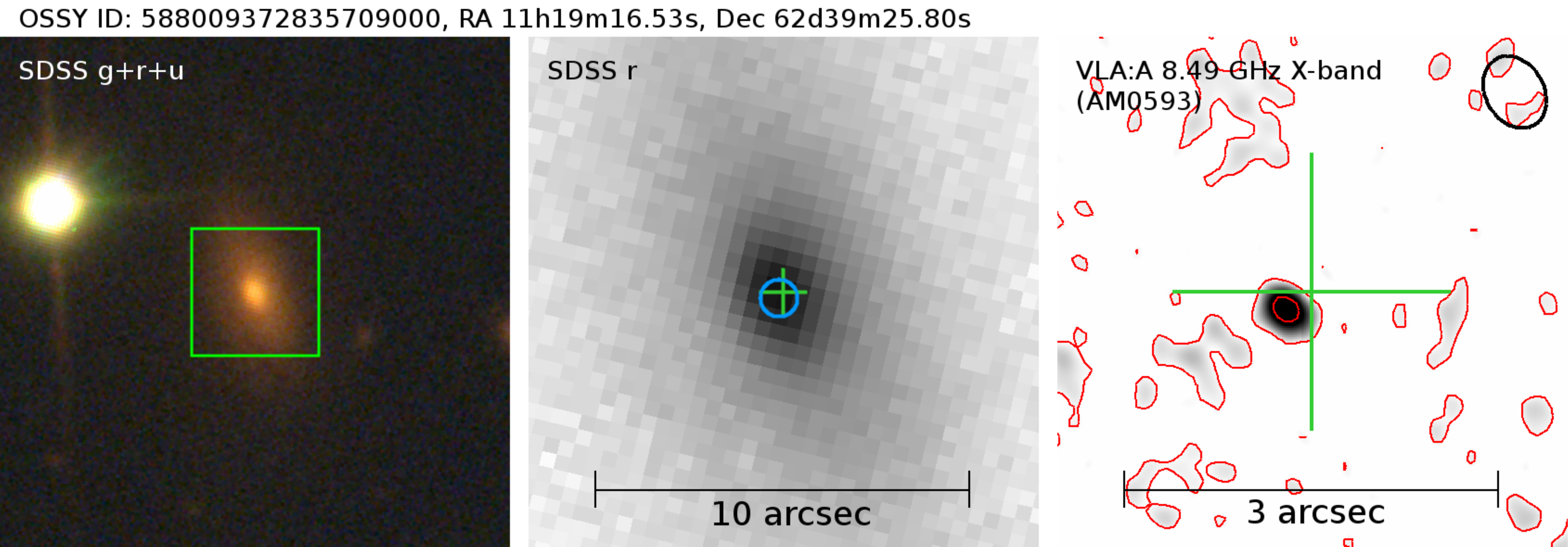}
\end{minipage}

\subsection{SDSS~J112039.94+504938.2}

\begin{flushleft}
AGN class: Seyfert, Morphology: compact, E (NED)
\end{flushleft}
Single CLASS detection of 13.8~mJy (NVSS~J112040+504936), offset by 159~mas from the SDSS position. The SDSS spectrum appears to be that of a passive elliptical galaxy, but the OSSY line strengths indicate that this galaxy is a Seyfert.

\vspace{0.2cm}

\noindent\begin{minipage}{0.48\textwidth}
	\centering
	\includegraphics[width=85mm]{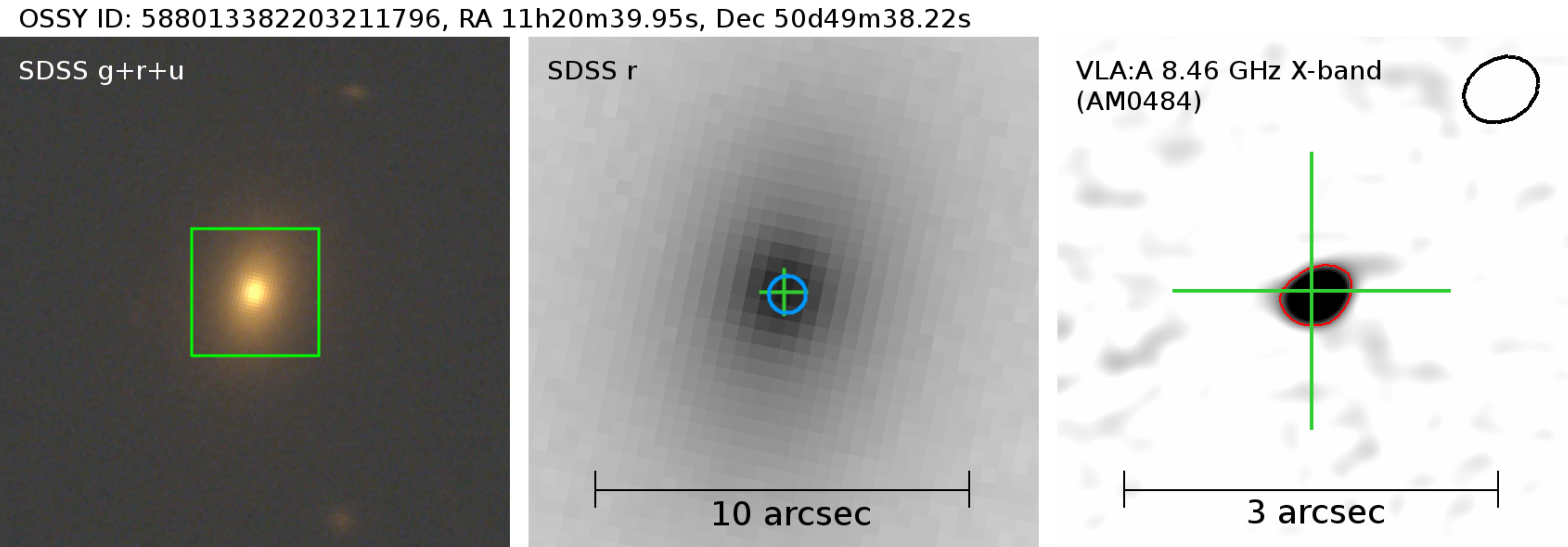}
\end{minipage}

\subsection{SDSS~J115000.07+552821.3}

\begin{flushleft}
AGN class: Seyfert
\end{flushleft}
Single CLASS detection of 70.0~mJy (NVSS~J115000+552821), offset by 1.13~arcsec from the SDSS position. Host galaxy shows signs of being disturbed, and appears a good candidate for an offset-AGN host. The offset of the radio source in SDSS~J115000.07+552821.3 has also been noted by \cite{DeVries2009}. This offset deviates from the fitted Rayleigh distribution with 4-sigma confidence.

\vspace{0.2cm}

\noindent\begin{minipage}{0.48\textwidth}
	\centering
	\includegraphics[width=85mm]{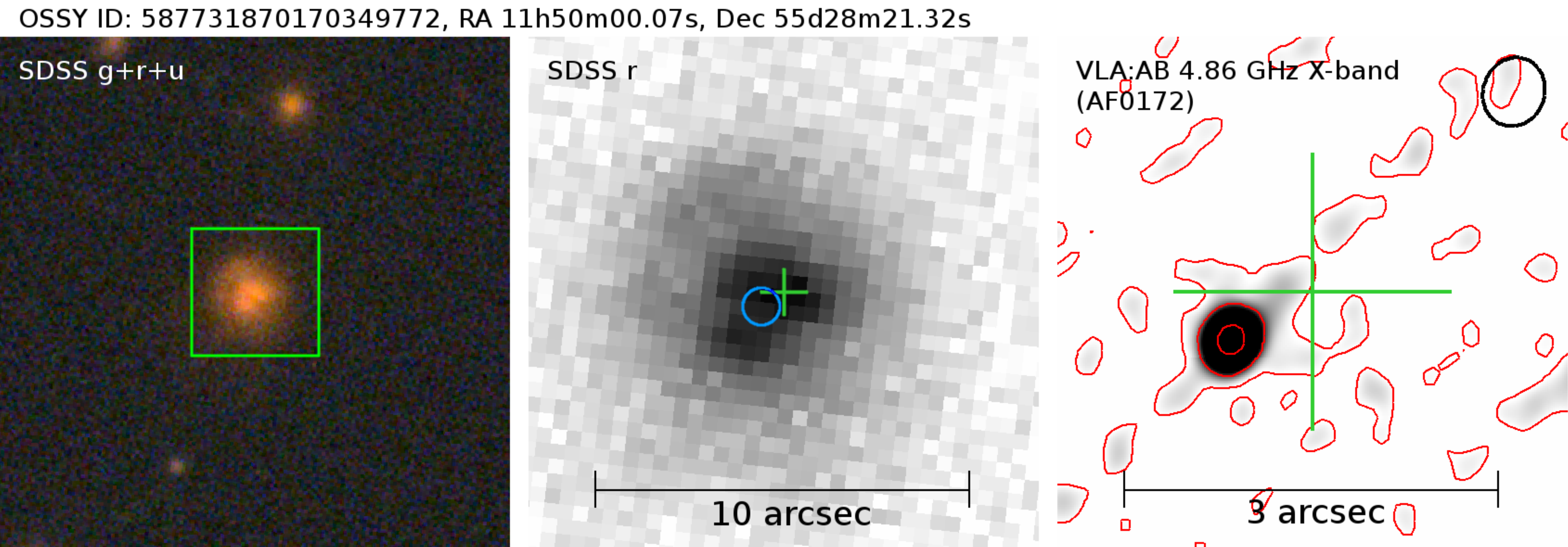}
\end{minipage}

\subsection{SDSS~J115133.06+050605.2}

\begin{flushleft}
Morphology: E, compact (NED)
\end{flushleft}
Single CLASS detection of 27.6~mJy (NVSS~J115133+050608), offset by 157~mas from the SDSS position. Optical images show an object about 4 arcsec to the north east. SDSS Spectrum of SDSS~J115133.06+050605.2 indicates a passive elliptical galaxy, although the OSSY line strengths suggest that this galaxy hosts a LINER.

\vspace{0.2cm}

\noindent\begin{minipage}{0.48\textwidth}
	\centering
	\includegraphics[width=85mm]{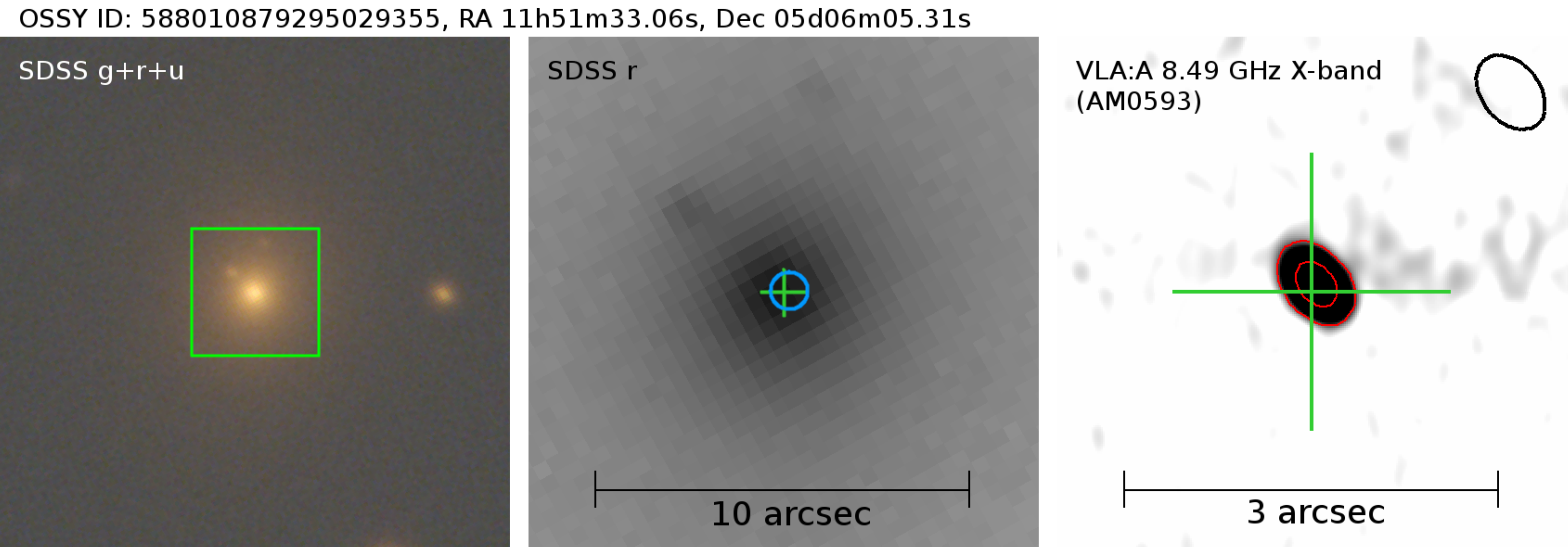}
\end{minipage}

\subsection{SDSS~J115410.41+122509.7}

Single CLASS detection of 187.5~mJy (NVSS~J115410+122509), offset by 221~mas from the SDSS position. The spectrum shows some strong narrow lines, especially [O\,\textsc{ii}]. This galaxy represents a very promising candidate for an offset-AGN host.

\vspace{0.2cm}

\noindent\begin{minipage}{0.48\textwidth}
	\centering
	\includegraphics[width=85mm]{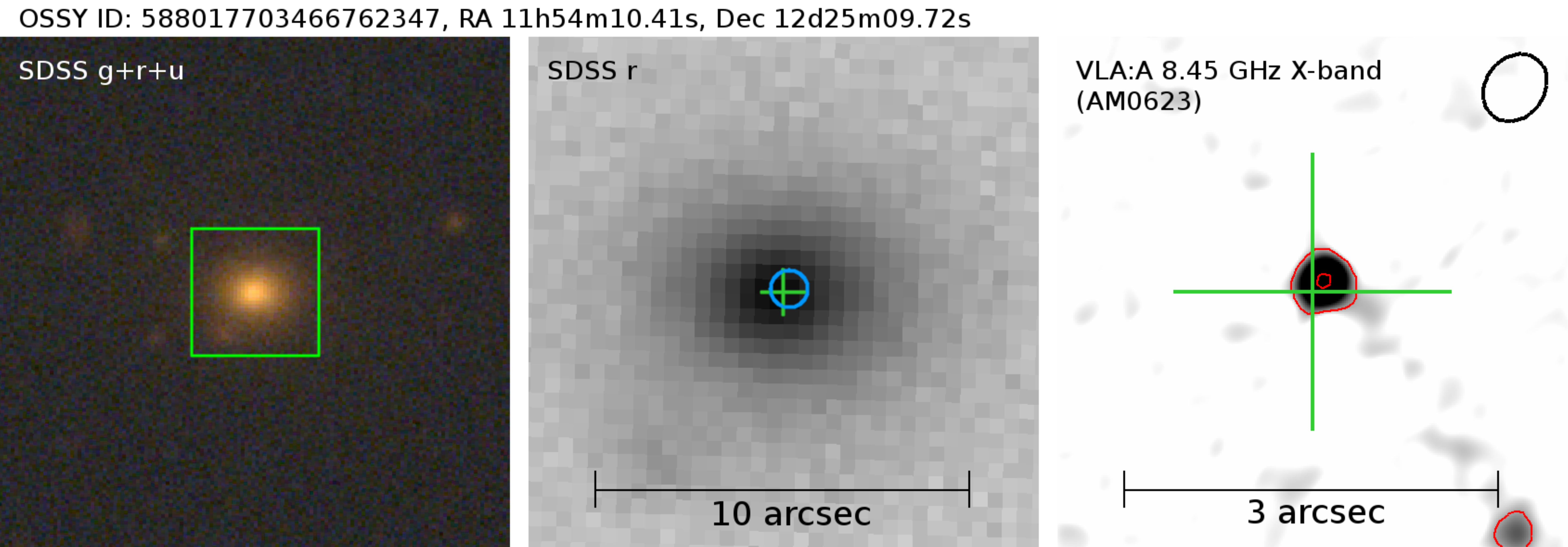}
\end{minipage}

\subsection{SDSS~J115905.67+582035.5 (MCG~+10-17-121)}

Single CLASS detection of 9.5~mJy (CRATES~J1159+5820), offset by 291~mas from the SDSS position. Compact \textit{Chandra} source detected at the same position. Galaxy shows some evidence of being disturbed. Radio imaging of this galaxy shows that the CLASS-detected source is the core of an extended radio galaxy \citep{KozielWierzbowska2012}. SDSS image shows a disturbed, possibly spiral, structure, which is unusual for the host of a radio galaxy. This offset deviates from the fitted Rayleigh distribution with 4-sigma confidence.

\vspace{0.2cm}

\noindent\begin{minipage}{0.48\textwidth}
	\centering
	\includegraphics[width=56mm]{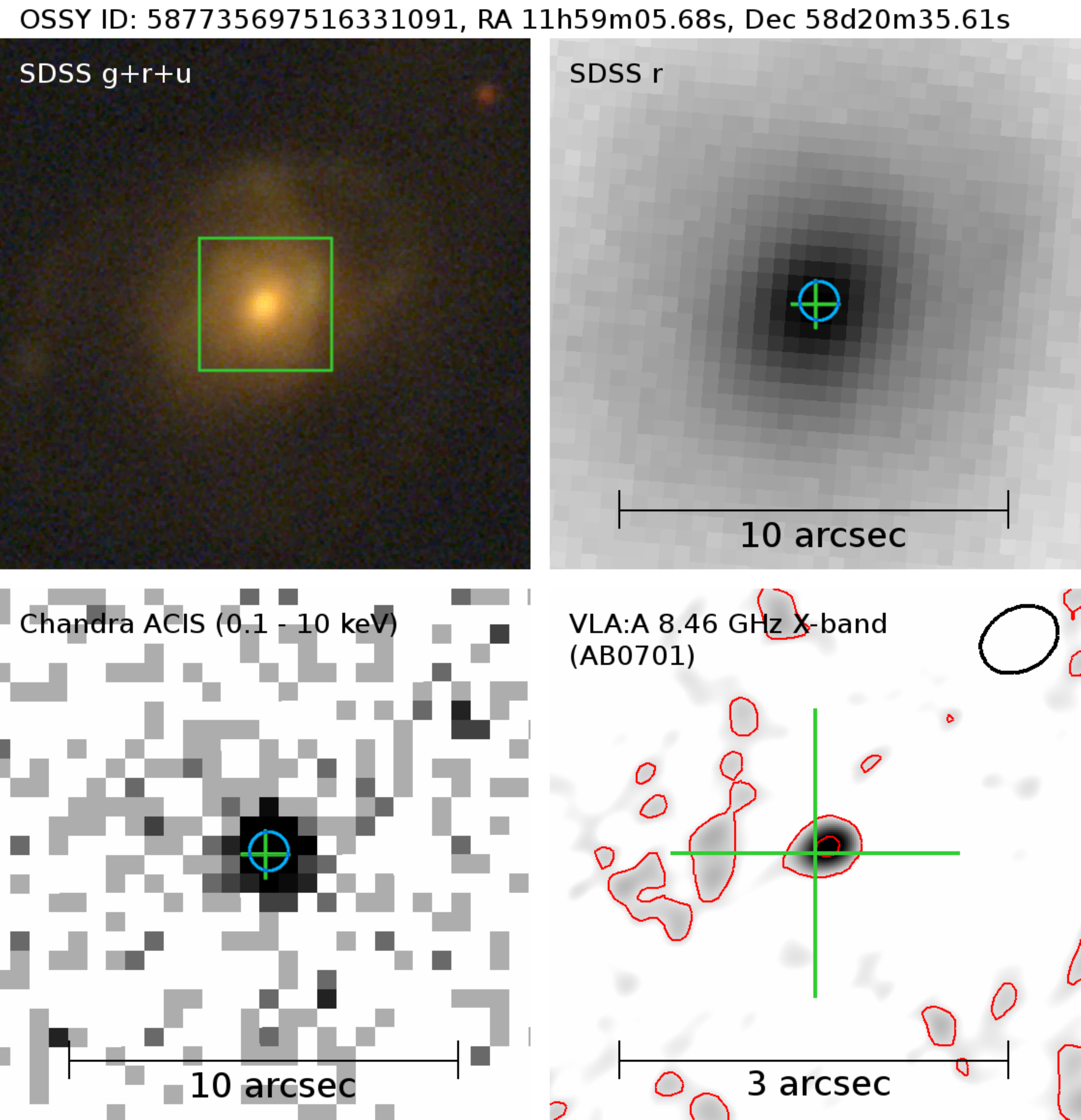}
	\begin{flushright}
	Chandra obs: 17116
	\end{flushright}
\end{minipage}

\subsection{SDSS~J122209.28+581421.5}

Single CLASS detection of 35.0~mJy (NVSS~J122209+581421), offset by 178~mas from the SDSS position. The SDSS spectrum shows little evidence of AGN activity, but the OSSY line strengths indicate that this source may be a LINER.

\vspace{0.2cm}

\noindent\begin{minipage}{0.48\textwidth}
	\centering
	\includegraphics[width=85mm]{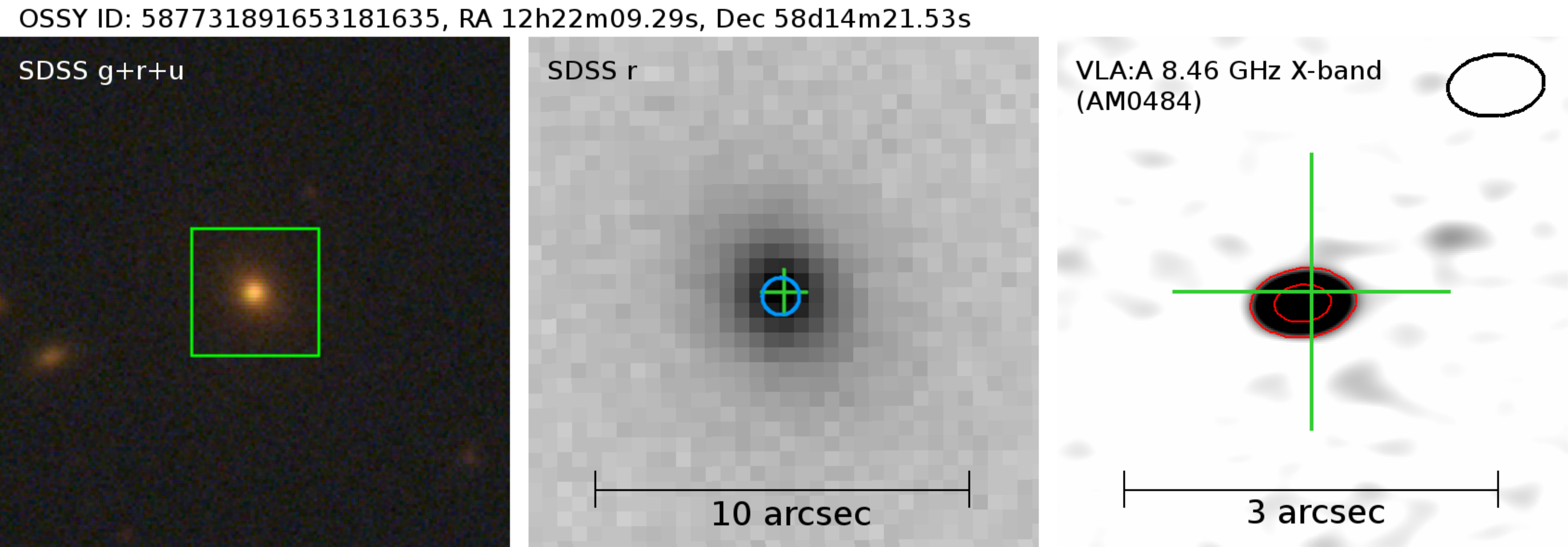}
\end{minipage}

\subsection{SDSS~J122513.09+321401.5}

Single CLASS detection of 80.5~mJy (NVSS~J122513+321401), offset by 408~mas from the SDSS position. Host galaxy is a large early-type galaxy at the centre of a cluster. The AGN is classified as a Seyfert~2 by \cite{Veron-Cetty2006}, but the OSSY line strengths suggest that this galaxy may be a LINER. Appears a good candidate for an offset-AGN host, with an offset that deviates from the fitted Rayleigh distribution with 5-sigma confidence.

\vspace{0.2cm}

\noindent\begin{minipage}{0.48\textwidth}
	\centering
	\includegraphics[width=85mm]{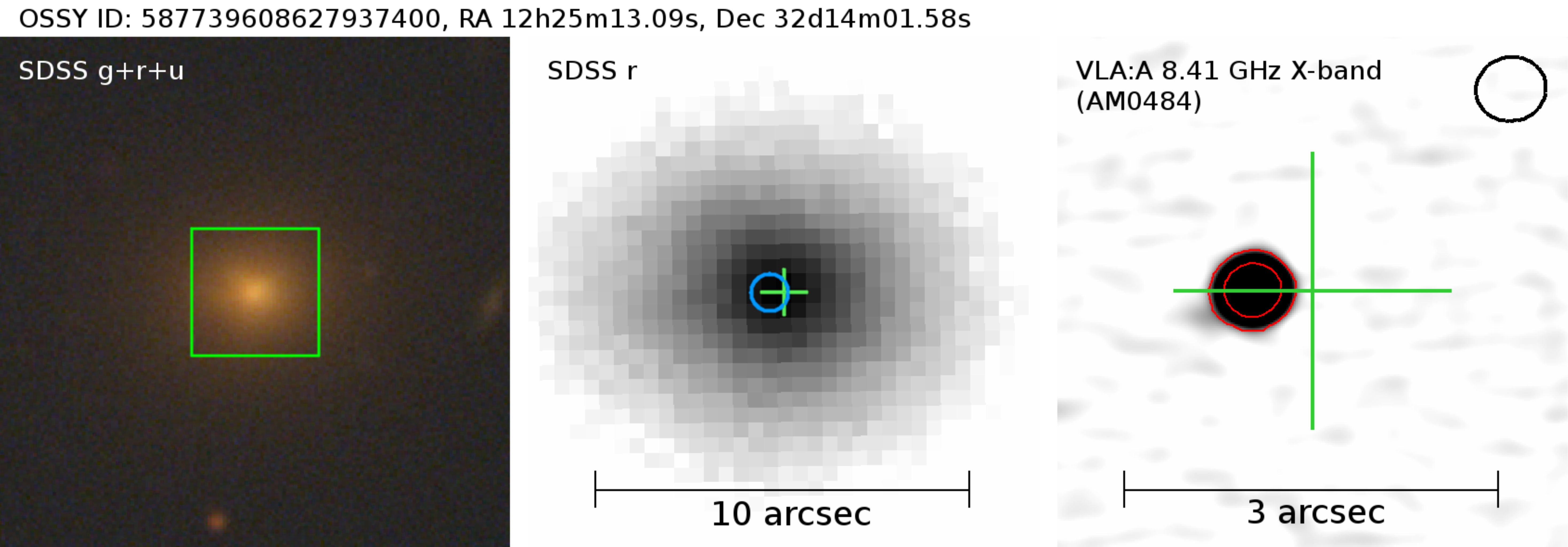}
\end{minipage}

\subsection{SDSS~J122622.47+640622.0}

Single CLASS detection of 70.7~mJy (CRATES~J122622.51+640622.0), offset by 185~mas from the SDSS position. The SDSS spectrum appears to be that of a passive elliptical galaxy, but the OSSY line strengths indicate that this galaxy may be a LINER.

\vspace{0.2cm}

\noindent\begin{minipage}{0.48\textwidth}
	\centering
	\includegraphics[width=85mm]{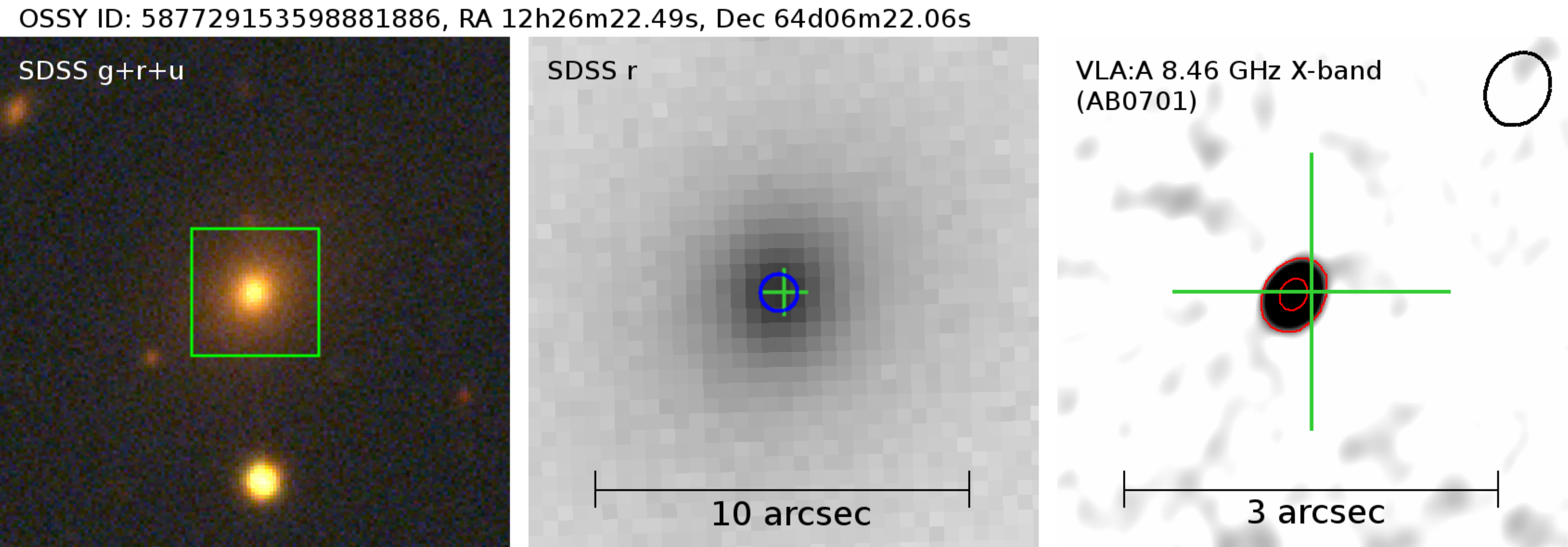}
\end{minipage}

\subsection{SDSS~J124135.08+285036.5}

Single CLASS detection of 31.4~mJy (NVSS~J124135+285034), offset at 467~mas from the SDSS position. Galaxy is extended along a north-west to south-east axis, and the offset of the radio source is towards the south east. The nucleus appears elongated, and we can't exclude the possibility of a double nucleus. The spectrum shows extremely strong [O\,\textsc{ii}] emission. This object was selected by \cite{Comerford2014} as an offset-AGN candidate on the basis of a spectral-line offset. This offset deviates from the fitted Rayleigh distribution with 5-sigma confidence, and the galaxy appears a good candidate for an offset-AGN host.

\vspace{0.2cm}

\noindent\begin{minipage}{0.48\textwidth}
	\centering
	\includegraphics[width=85mm]{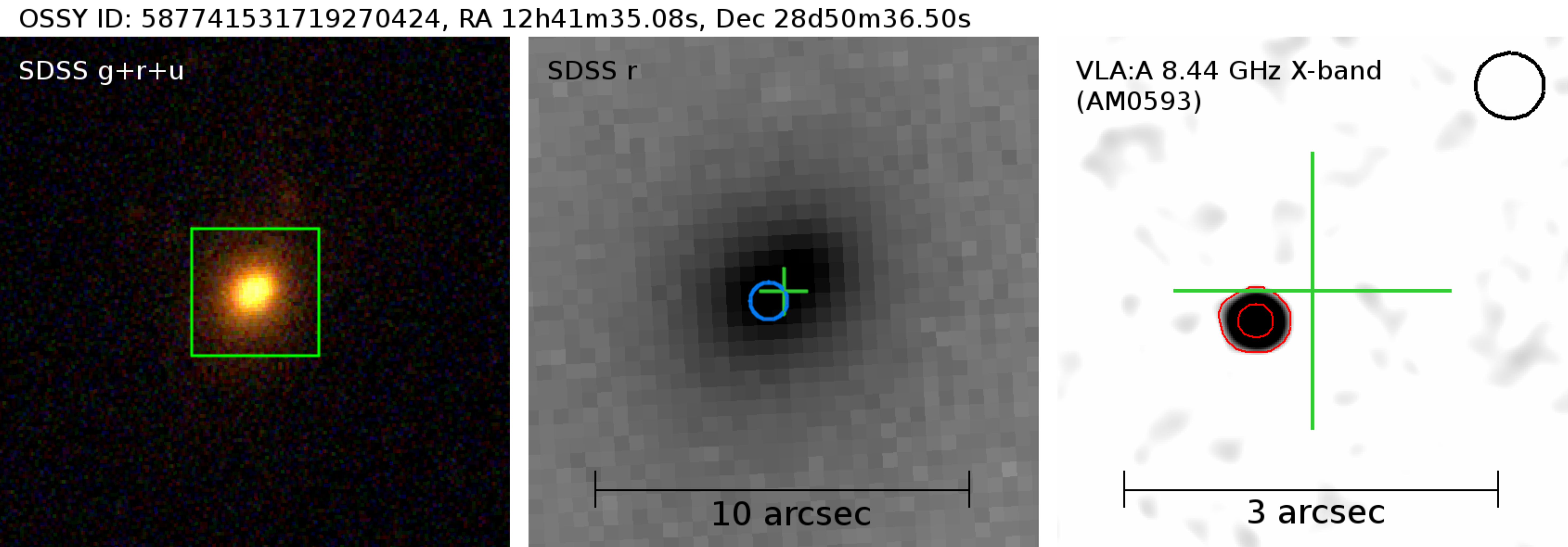}
\end{minipage}

\subsection{SDSS~J125433.25+185602.1}

Single CLASS detection of 128.2~mJy (NVSS~J125433+185602), offset by 273~mas from the SDSS position. This source lies within Abell 1638, and \cite{Owen1993} show a 20~cm VLA map with extended structure. This source has a dust lane, which would make the optical position less reliable, but the offset deviates from the fitted Rayleigh distribution with 4-sigma confidence.

\vspace{0.2cm}

\noindent\begin{minipage}{0.48\textwidth}
	\centering
	\includegraphics[width=85mm]{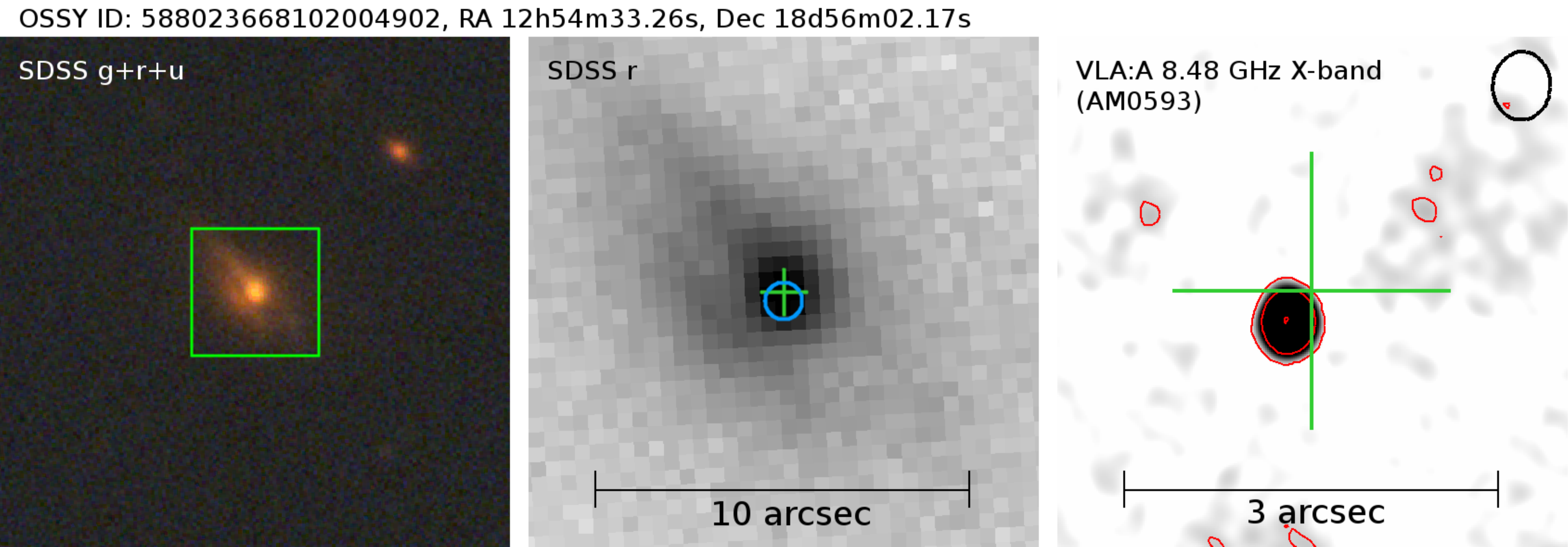}
\end{minipage}

\subsection{SDSS~J131503.51+243707.7 (IC~860, MCG~+04-31-015)}

\begin{flushleft}
Morphology: S? (NED)
\end{flushleft}
Two CLASS detections, with the brightest (10.3~mJy) offset from the SDSS position by 160~mas. IC~860 has been found to be somewhat unusual \citep{Kazes1988}, with behaviour described as typical of both an interacting and a starburst galaxy. The weak \textit{Chandra} detection (\citealt{Lehmer2010} placed an upper limit of ${\rm log} L_{\rm 2-10\,keV} = {\rm 40.19~erg~s^{-1}}$) fails to support the presence of an AGN.

\vspace{0.2cm}

\noindent\begin{minipage}{0.48\textwidth}
	\centering
	\includegraphics[width=56mm]{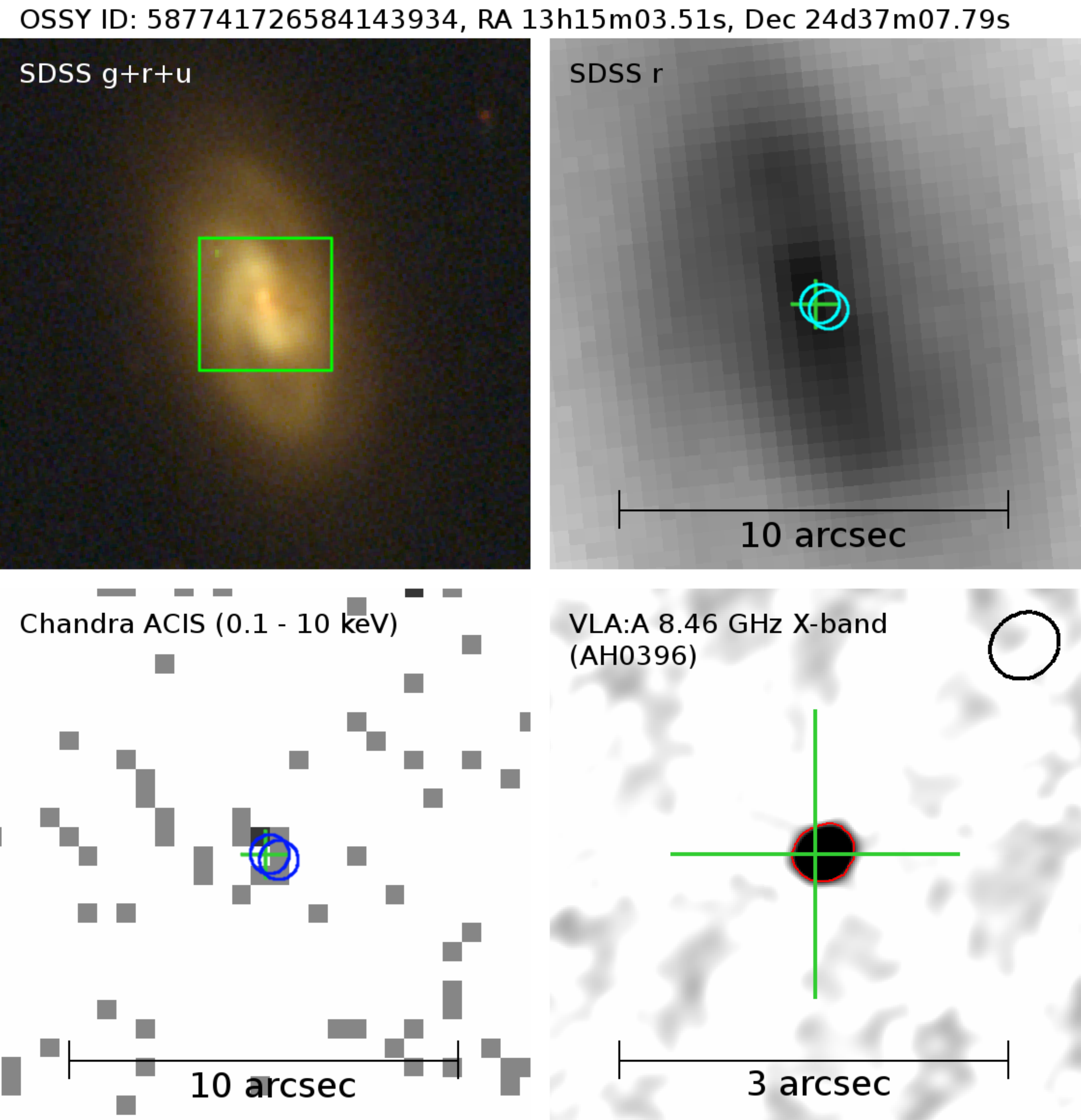}
	\begin{flushright}
	Chandra obs: 10400
	\end{flushright}
\end{minipage}

\subsection{SDSS~J133435.06+344639.9 (NGC~5228; UGC~8556; MCG~+06-30-043)}

\begin{flushleft}
AGN class: Seyfert, Morphology: S0?, S0/E (NED)
\end{flushleft}
Single CLASS detection of 20.9~mJy (NVSS~J133435+344639), offset by 176~mas from the SDSS position. Possible dust cloud, dust lane, or other form of obscuration, to the west and north. The SDSS spectrum appears to be that of a passive elliptical galaxy, but the OSSY line strengths indicate that this galaxy is a Seyfert.

\vspace{0.2cm}

\noindent\begin{minipage}{0.48\textwidth}
	\centering
	\includegraphics[width=85mm]{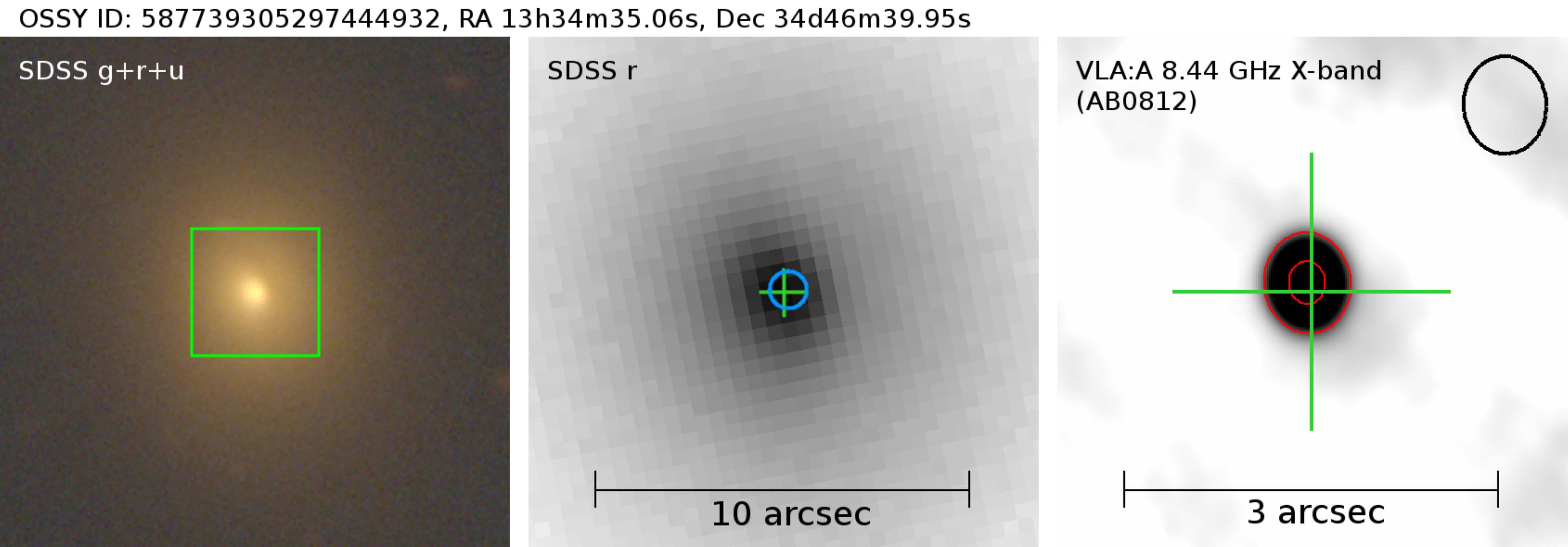}
\end{minipage}

\subsection{SDSS~J133621.18+031951.0 (MCG~+01-35-014)}

\begin{flushleft}
Morphology: E (NED)
\end{flushleft}
Brightest CLASS detection has a flux density of 21.5~mJy (NVSS~J133621+031952; offset by 162~mas from the SDSS position). An extremely faint CLASS source is found nearby. The SDSS spectrum appears to be that of a passive elliptical galaxy, but the OSSY line strengths indicate that this galaxy may be a LINER.

\vspace{0.2cm}

\noindent\begin{minipage}{0.48\textwidth}
	\centering
	\includegraphics[width=85mm]{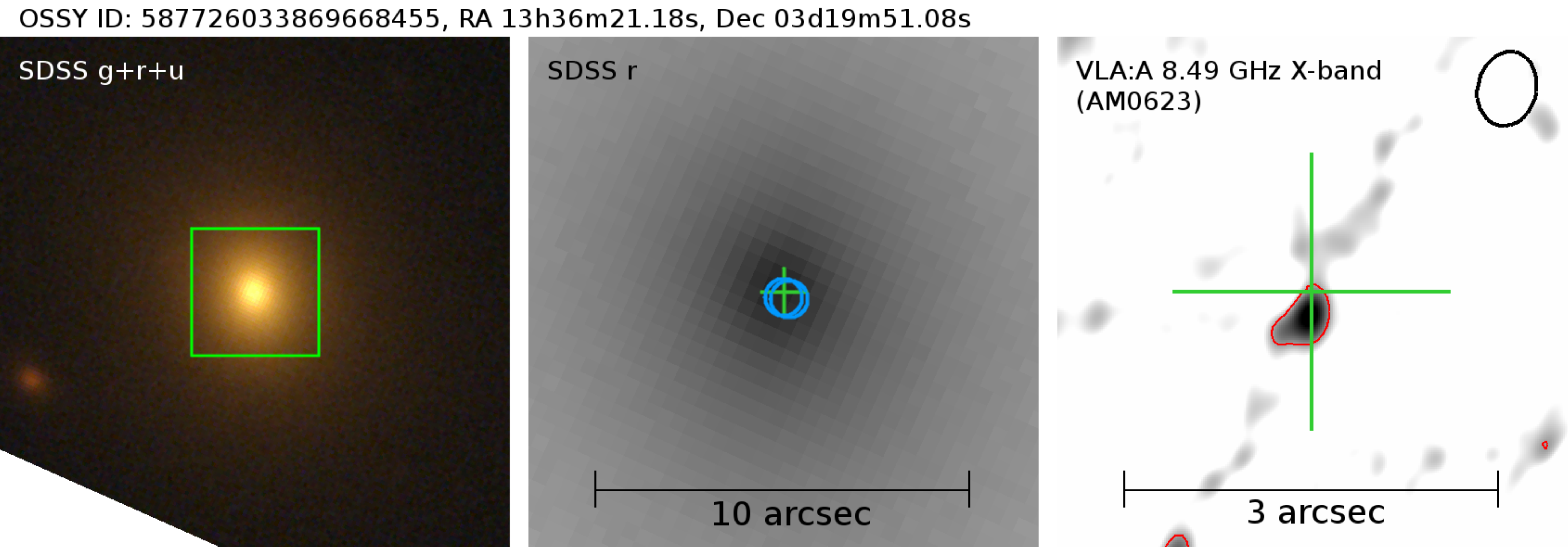}
\end{minipage}

\subsection{SDSS~J134243.62+050432.1}

Bright CLASS detection of 168.7~mJy (NVSS~J134243+050431, 4C +05.57), offset by 163~mas from the SDSS position. A second bright source (51.2~mJy) is offset by 374~mas, and three fainter sources of up to 13.7~mJy in flux density are offset by at least 333~mas. SDSS spectrum appears to be a typical narrow-line AGN. The AGN is classified as a Seyfert~1 by \cite{Veron-Cetty2006}, but the OSSY line strengths indicate that this galaxy may be a LINER. This galaxy appears a good candidate for an offset-AGN host.

\vspace{0.2cm}

\noindent\begin{minipage}{0.48\textwidth}
	\centering
	\includegraphics[width=85mm]{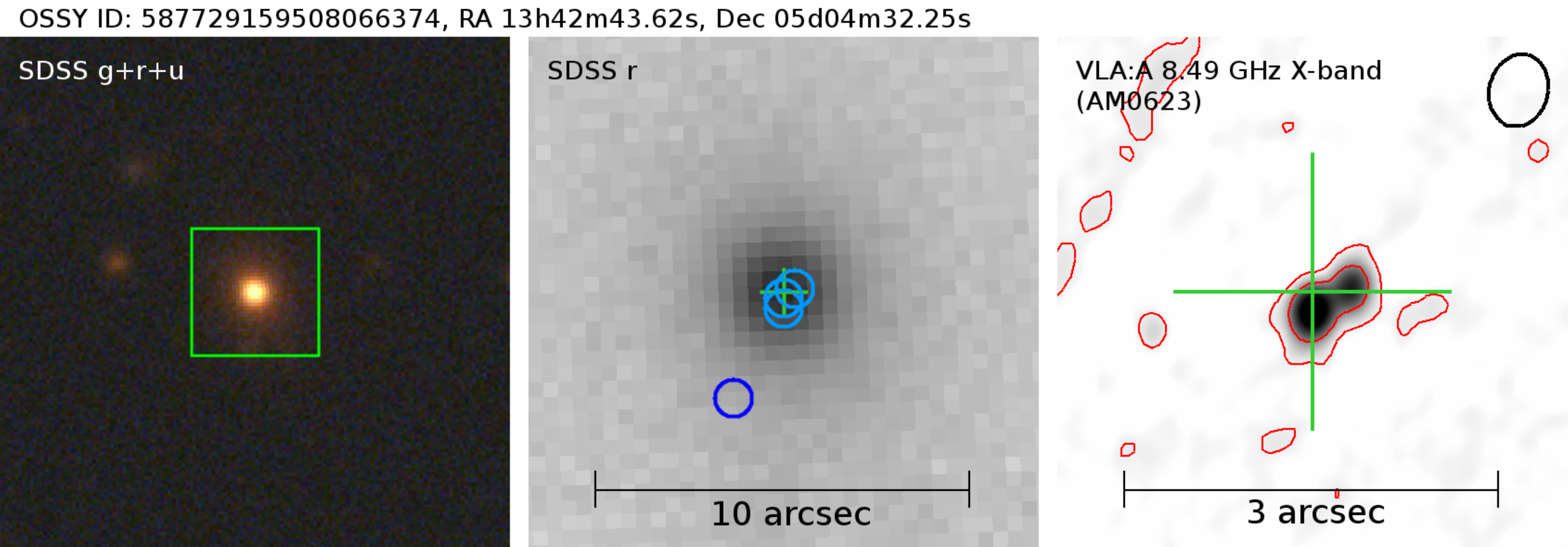}
\end{minipage}

\subsection{SDSS~J134442.15+555313.8 (Mrk~273; UGC~8696; MCG~+09-23-004)}

\begin{flushleft}
AGN class: Seyfert, Morphology: pec (NED)
\end{flushleft}
Brightest and nearest CLASS detection of 33.6~mJy (CRATES~J134442.13+555313.5), offset by 559~mas from the SDSS position. There is a fainter 4.0~mJy detection slightly further away. The AGN is classified as a Seyfert~2 by \cite{Khachikian1974}. The \textit{Chandra} detection is offset from the optical centre of the galaxy, and in the same direction as the two CLASS detections. It has been suggested \citep{U2013} that this galaxy hosts a dual AGN system. The offset deviates from the fitted Rayleigh distribution with 5-sigma confidence.

\vspace{0.2cm}

\noindent\begin{minipage}{0.48\textwidth}
	\centering
	\includegraphics[width=56mm]{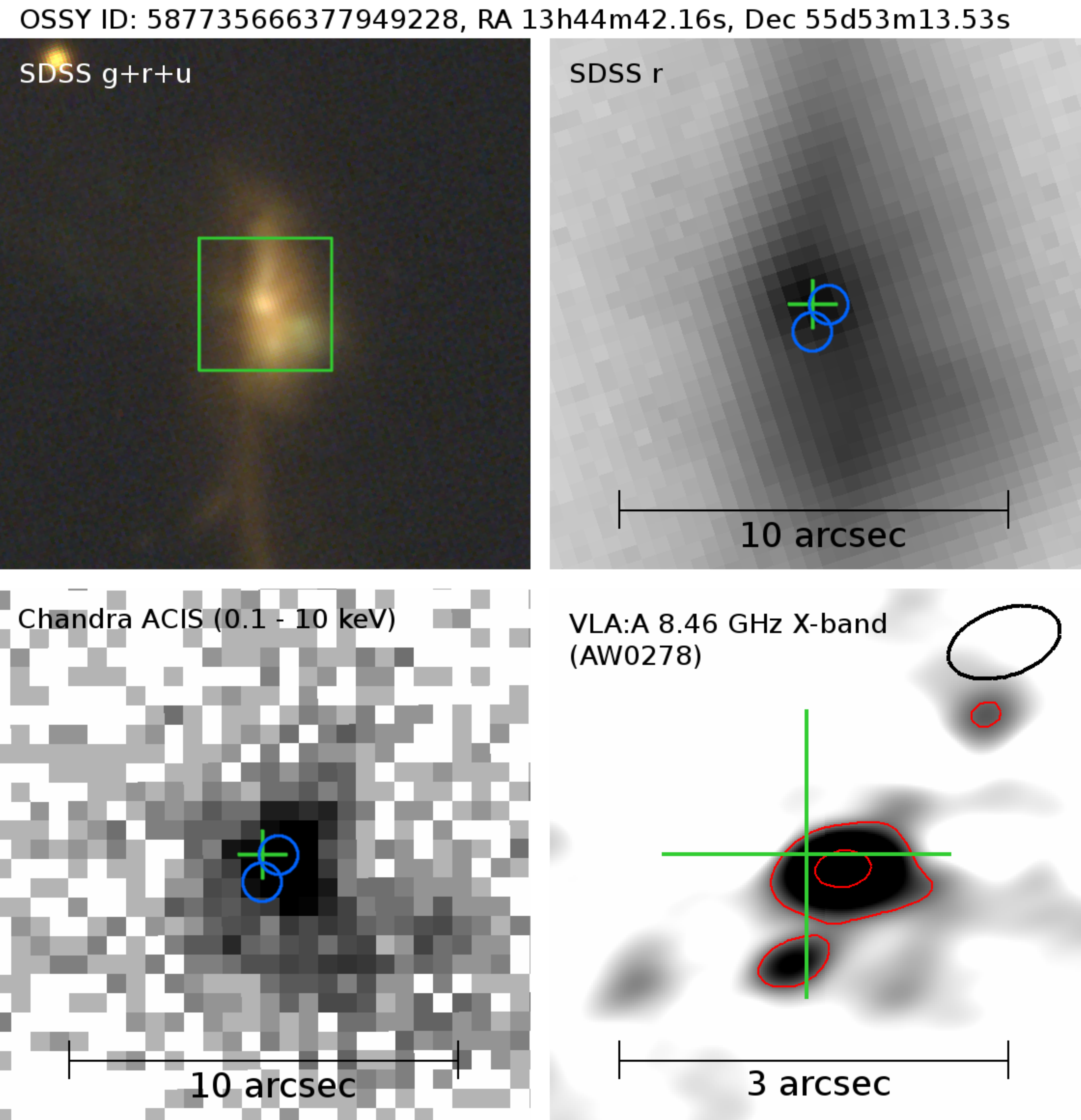}
	\begin{flushright}
	Chandra obs: 809
	\end{flushright}
\end{minipage}

\subsection{SDSS~J135022.12+094010.7}

Single CLASS detection of 250.4~mJy (NVSS~J135022+094010), offset by 154~mas from the SDSS position. A compact \textit{Chandra} source is also detected at the same position. This source is found within the VLBA calibrator list, with milliarcsec-precision position. The VLBA/SDSS offset is a slightly greater 166~mas. The AGN is classified as a Seyfert~1.9 by \cite{Veron-Cetty2006}, but the OSSY line strengths indicate that this galaxy may be a LINER. The SDSS images suggest a lot of optical sub-structure, and the spectrum looks like a typical narrow-line AGN.

\vspace{0.2cm}

\noindent\begin{minipage}{0.48\textwidth}
	\centering
	\includegraphics[width=56mm]{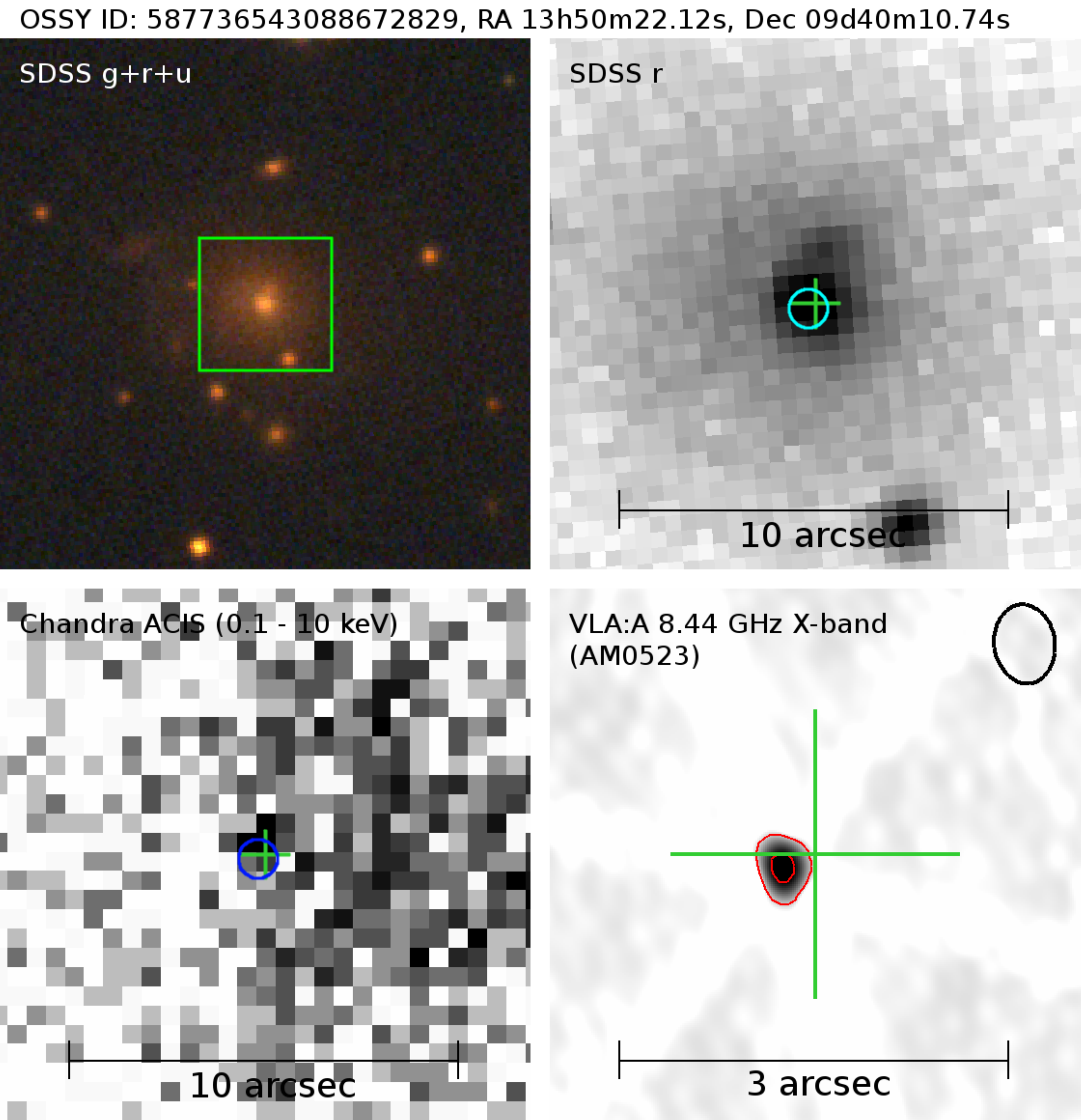}
	\begin{flushright}
	Chandra obs: 14021
	\end{flushright}
\end{minipage}

\subsection{SDSS~J135036.01+334217.3 (NGC~5318~NED01; UGC~8751)}

\begin{flushleft}
Morphology: SO? (NED)
\end{flushleft}
Single CLASS detection of 95.8~mJy (NVSS~J135036+334218), offset by 208~mas from the SDSS position. NGC~5318 is the brightest member in a group of galaxies, and has been studied as an example of a spiral galaxy with a bright nuclear radio source \citep[see][and references therein]{Mirabel1983}. The AGN is classified as a possible Seyfert by \cite{Veron-Cetty2006}, but the OSSY line strengths indicate that this galaxy may be a LINER.

\vspace{0.2cm}

\noindent\begin{minipage}{0.48\textwidth}
	\centering
	\includegraphics[width=85mm]{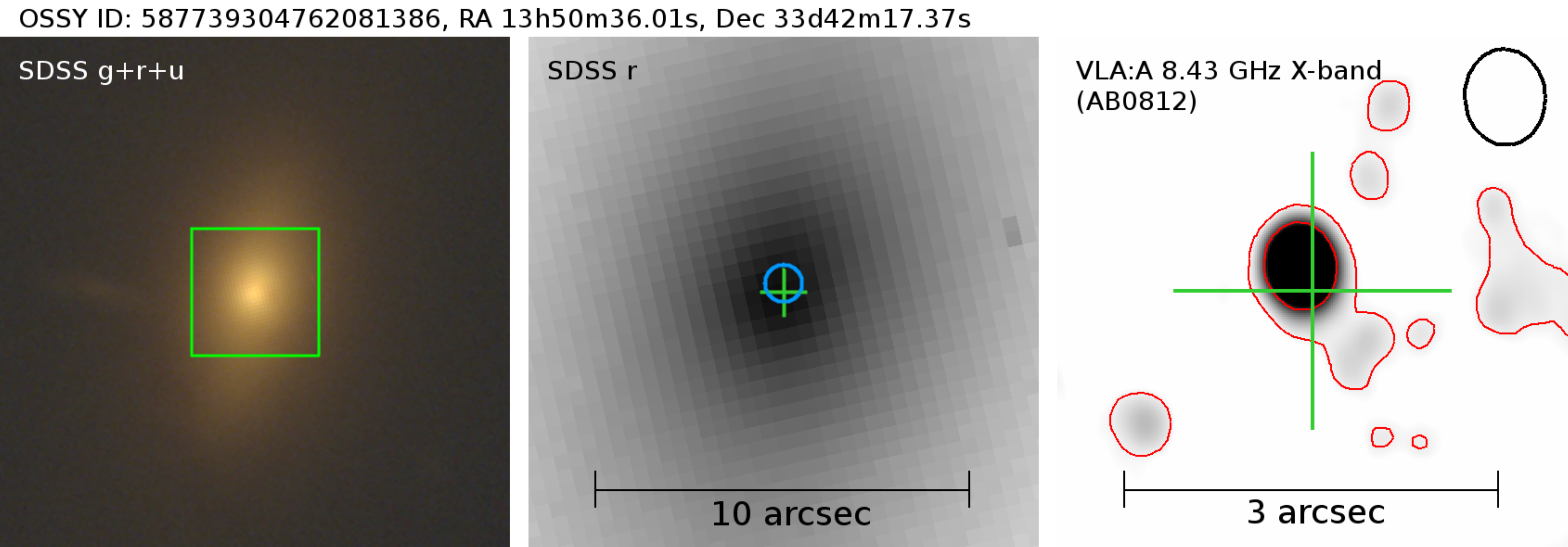}
\end{minipage}

\subsection{SDSS~J135927.60+465045.9}

Single CLASS detection of 10.2~mJy, offset by 249~mas from the SDSS position. SDSS spectrum appears to be that of a passive elliptical, but optical images hint that a faint disk may be present, and the OSSY line strengths indicate that this galaxy is a LINER. This source, albeit a little weak, appears to be a good candidate for an offset AGN.

\vspace{0.2cm}

\noindent\begin{minipage}{0.48\textwidth}
	\centering
	\includegraphics[width=85mm]{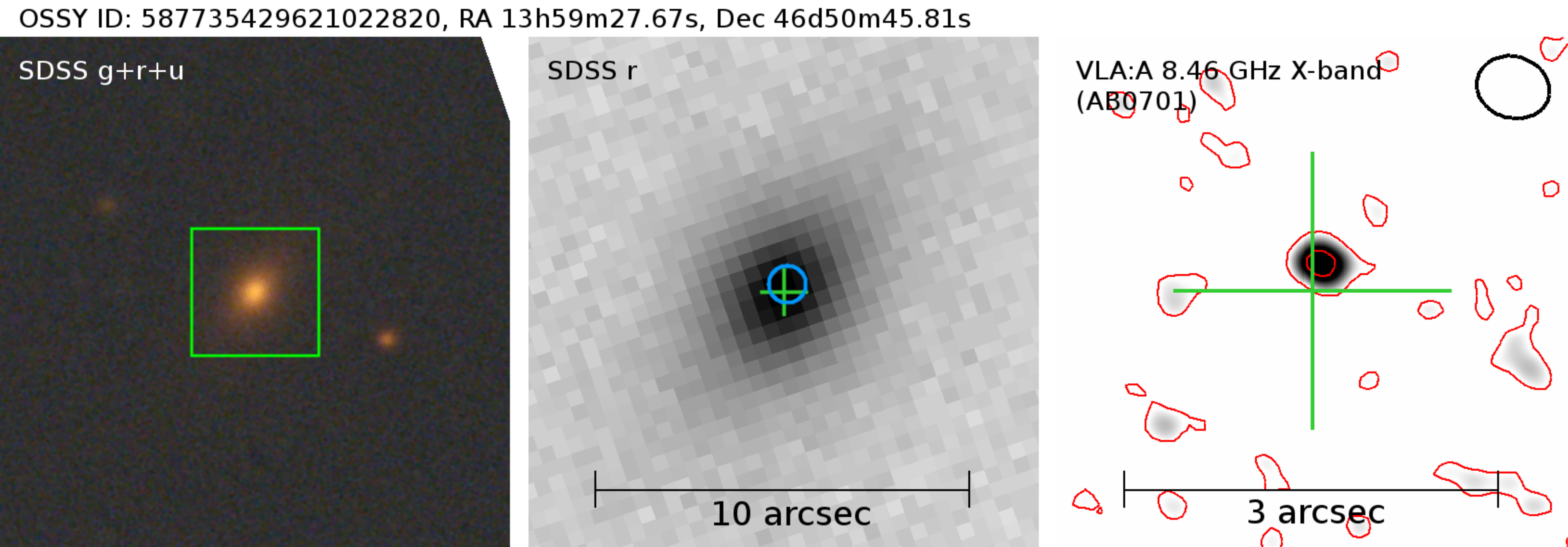}
\end{minipage}

\subsection{SDSS~J142730.26+540923.7}

Two CLASS sources detected in almost the same position (NVSS~J142730+540923), one of 19.5~mJy (offset by 173~mas) and another of 18.9~mJy (offset by 273~mas). These detections are likely to be from a single source, and have somehow become duplicated in the CLASS database. The AGN is classified as a possible BL~Lacertae object by \cite{Veron-Cetty2006}, and \cite{Nilsson2003} include this galaxy in their study of BL Lac hosts. However, the spectrum does not show an obviously depressed {4000~\AA} break as one would expect for a BL Lac. There is a possible face-on disk with faint spiral structure. The larger offset deviates from the fitted Rayleigh distribution with 4-sigma confidence.

\vspace{0.2cm}

\noindent\begin{minipage}{0.48\textwidth}
	\centering
	\includegraphics[width=85mm]{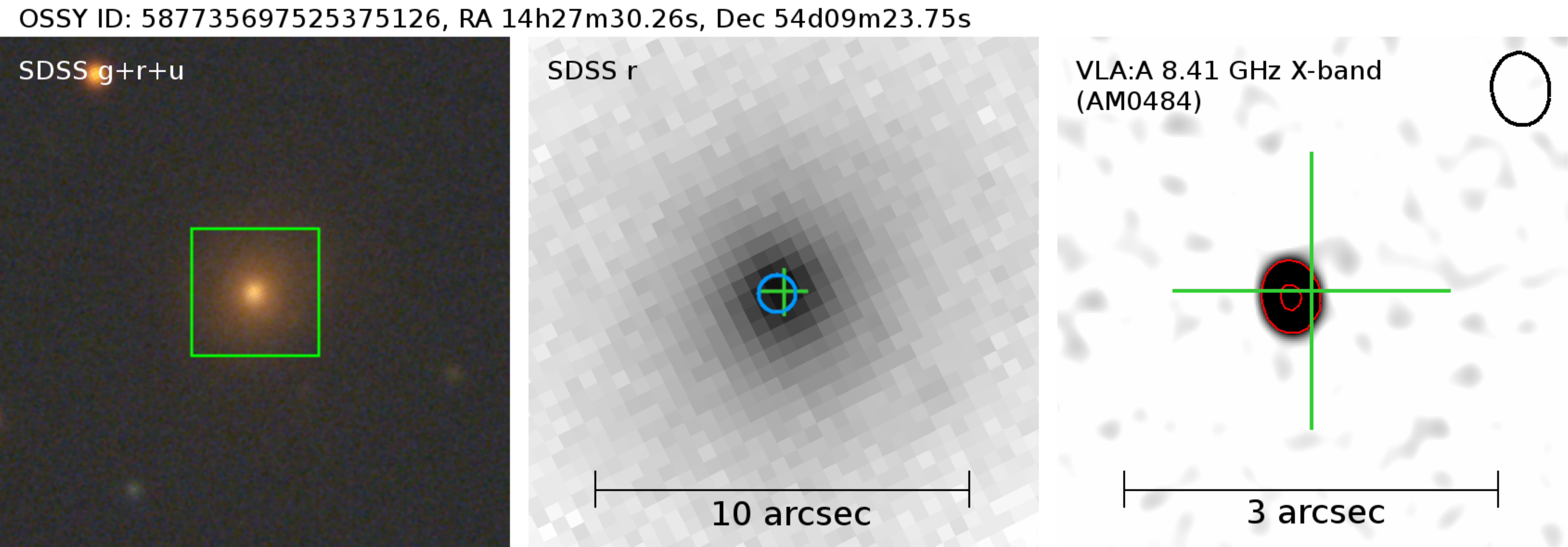}
\end{minipage}

\subsection{SDSS~J144441.09+142346.9}

Single CLASS detection of 22.9~mJy (NVSS~J144441+142347), offset by 177~mas from the SDSS position. The SDSS spectrum looks like that of a passive elliptical galaxy, but the OSSY line strengths indicate that this galaxy may be a LINER.

\vspace{0.2cm}

\noindent\begin{minipage}{0.48\textwidth}
	\centering
	\includegraphics[width=85mm]{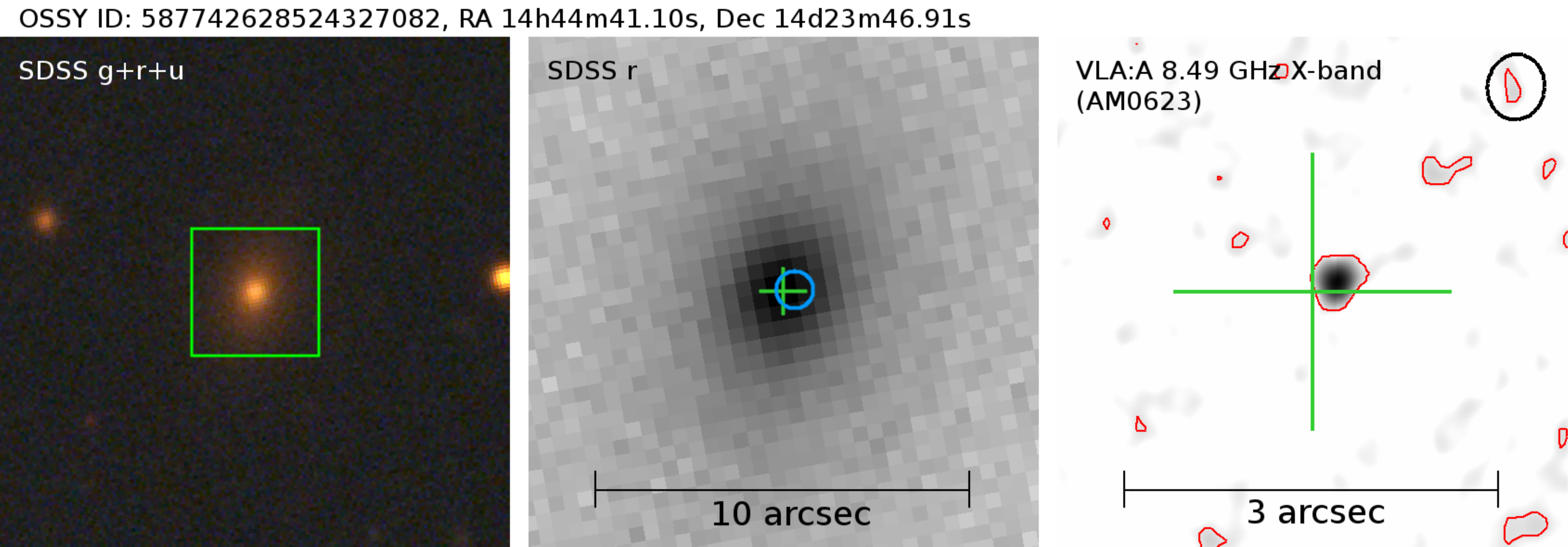}
\end{minipage}

\subsection{SDSS~J144607.04+090338.2}

Strong CLASS detection of 76.2~mJy (NVSS~J144607+090338; CRATES~J144607.03+090338.3), offset by 152~mas from the SDSS position. A very faint source is found further from the core. The SDSS spectrum looks like that of a passive galaxy, but the OSSY line strengths suggest that this galaxy may be a LINER.

\vspace{0.2cm}

\noindent\begin{minipage}{0.48\textwidth}
	\centering
	\includegraphics[width=85mm]{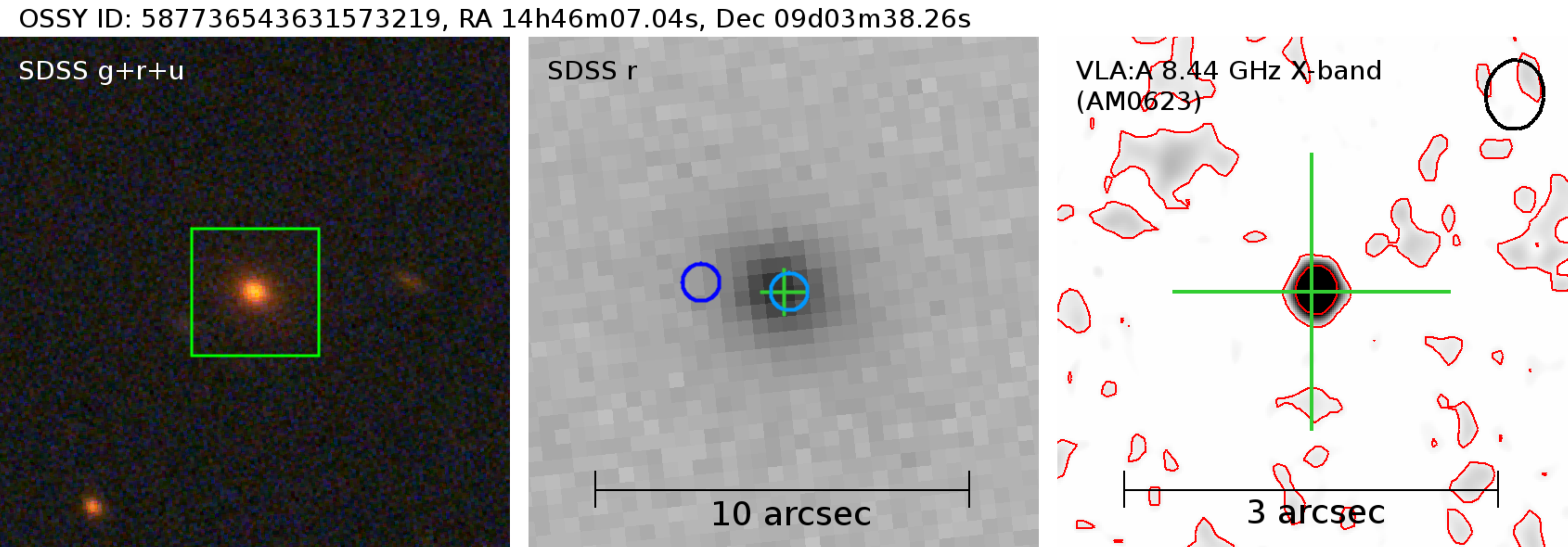}
\end{minipage}

\subsection{SDSS~J153457.21+233013.3 (IC~4553; Arp 220)}

\begin{flushleft}
Morphology: S?, pec (NED)
\end{flushleft}
Two bright CLASS detections of 60.1~mJy and 59.6~mJy, offset at 1.69 and 1.01~arcsec respectively (one or both of these sources are NVSS~J153457+233011) from the SDSS position. Several fainter sources of up to 13.6~mJy scattered at offsets of 0.91 and 3.60~arcsec. \textit{Chandra} detection is strong, but appears slightly resolved. The AGN is classified as a Seyfert by \cite{Veron-Cetty2006}, but the OSSY line strengths indicate that this galaxy may be a LINER. Arp 220 is a well-known merging galaxy \citep{Arp1966, Joseph1985, Casoli1988, Graham1990}. These offsets deviate from the fitted Rayleigh distribution with 5-sigma confidence.

\vspace{0.2cm}

\noindent\begin{minipage}{0.48\textwidth}
	\centering
	\includegraphics[width=56mm]{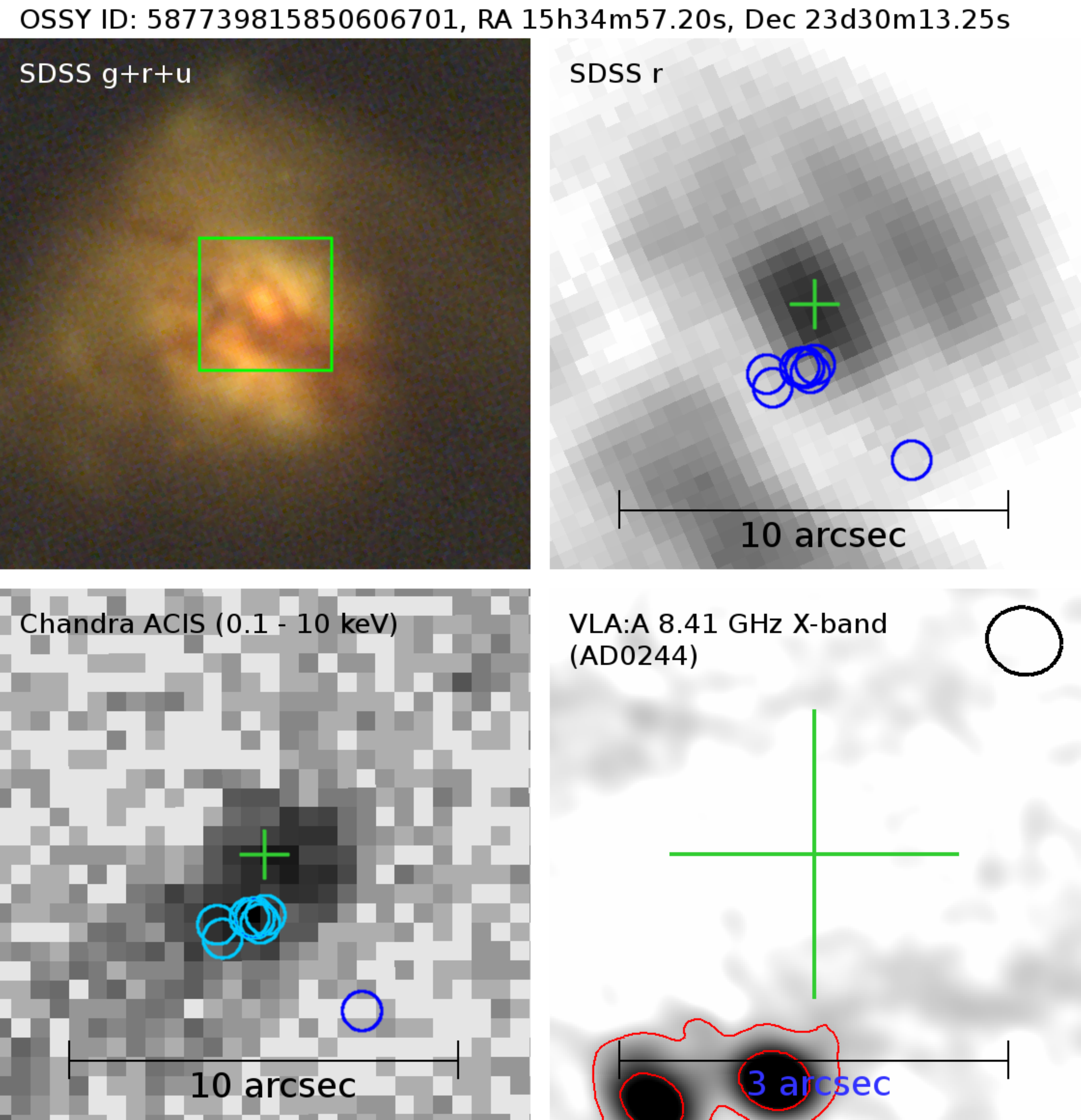}
	\begin{flushright}
	Chandra obs: 16092
	\end{flushright}
\end{minipage}

\subsection{SDSS~J154912.33+304716.4}

\begin{flushleft}
AGN class: Seyfert
\end{flushleft}
Nearest CLASS detection is 37.3~mJy (FIRST~J154912.3+304716), offset by 370~mas from the SDSS position. Two more bright sources, 38.0~mJy and 17.7~mJy, are offset by just over two arcsec. The 16.2~mJy source to the south-east (4C +30.29; NVSS~J154912+304715) is coincident with a compact \textit{Chandra} source, and with a redshift of 1.17 (NED) is probably a background quasar. This galaxy is a well-known gravitational lens system \citep{Lehar1993}, where a lobe of a background source is imaged into an Einstein ring. No X-ray sources are found in the nucleus of SDSS~J154912.33+304716.4, and this galaxy is not an offset-AGN host.

\vspace{0.2cm}

\noindent\begin{minipage}{0.48\textwidth}
	\centering
	\includegraphics[width=56mm]{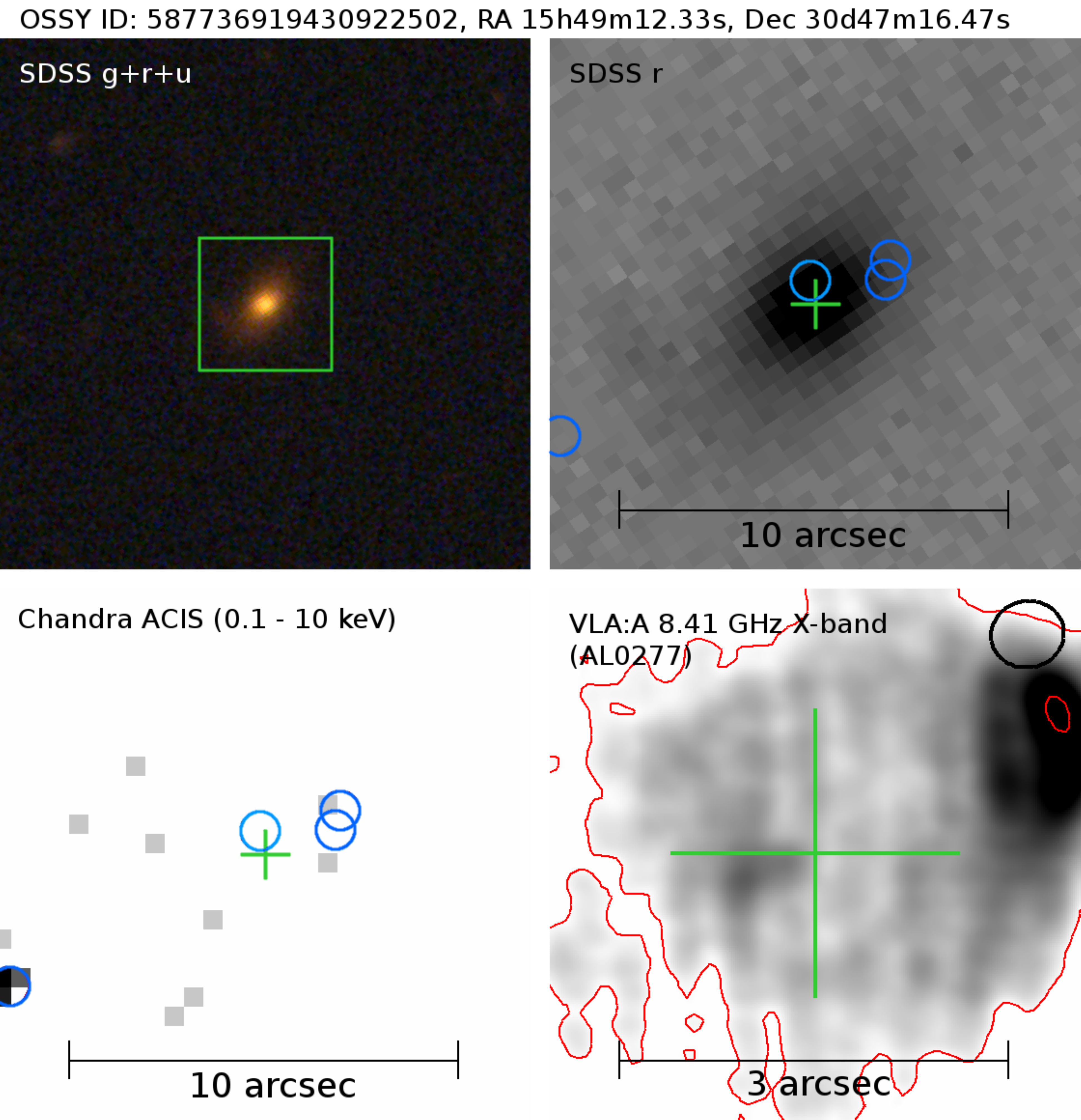}
	\begin{flushright}
	Chandra obs: 5716
	\end{flushright}
\end{minipage}

\subsection{SDSS~J162719.15+483126.7}

Single CLASS detection of 32.5~mJy (NVSS~J162719+483127), offset by 164~mas from the SDSS position. The spectrum looks like that of a mildly-active galaxy, and this galaxy is a good offset-AGN candidate.

\vspace{0.2cm}

\noindent\begin{minipage}{0.48\textwidth}
	\centering
	\includegraphics[width=85mm]{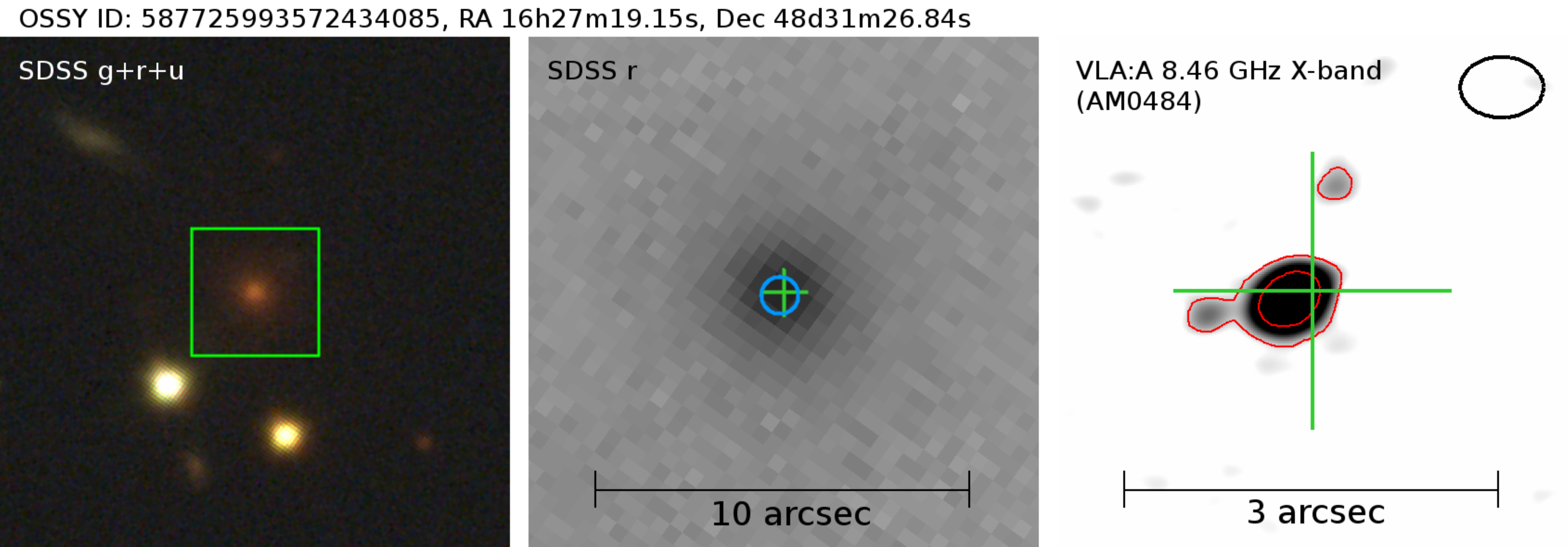}
\end{minipage}

\subsection{SDSS~J215527.22+120500.6}

\begin{flushleft}
AGN class: Seyfert
\end{flushleft}
Single CLASS detection of 58.8~mJy (NVSS~J215527+120501), offset by 248~mas from the SDSS position. The galaxy has a disk which is bluer than the nucleus, and the spectrum is that of a narrow-line radio galaxy.

\vspace{0.2cm}

\noindent\begin{minipage}{0.48\textwidth}
	\centering
	\includegraphics[width=85mm]{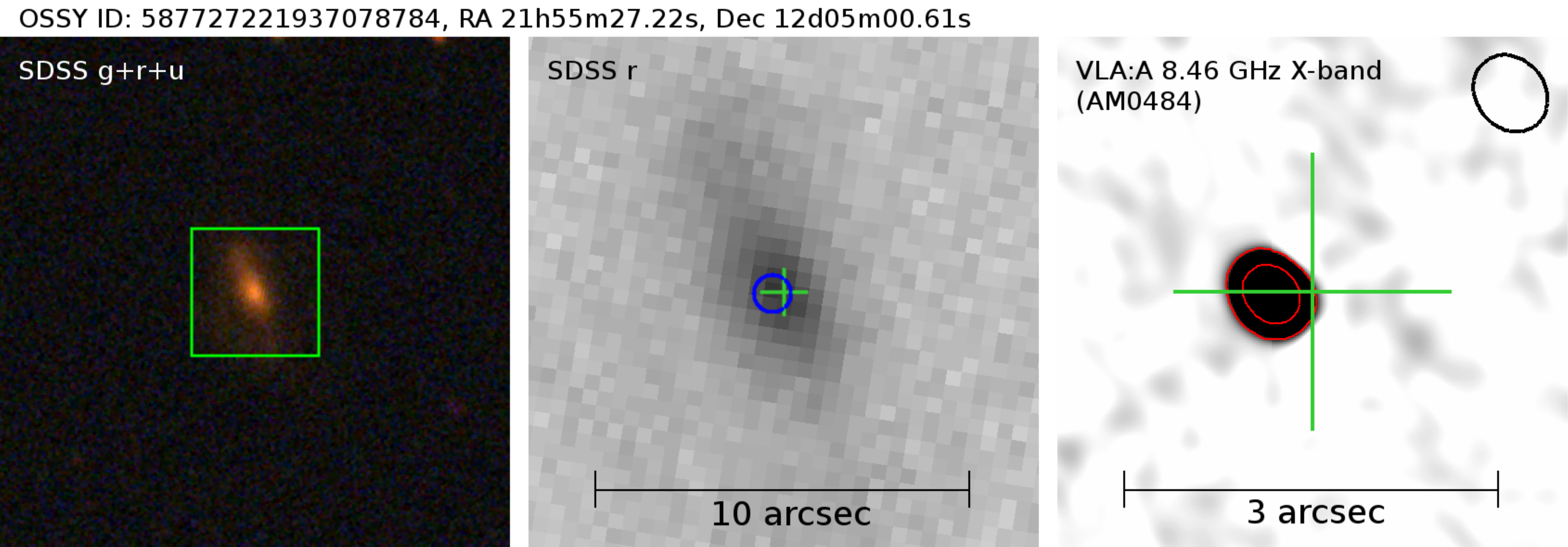}
\end{minipage}

\label{lastpage}
\end{document}